\documentclass[
aps,prd,
showpacs,twocolumn,notitlepage,
amssymb,amsmath,amsfonts,mathrsfs,
nofootinbib,superscriptaddress,
floats,floatfix
]{revtex4-2}

\usepackage{graphicx}
\usepackage{xcolor}
\usepackage[colorlinks=true]{hyperref}
\hypersetup{citecolor=cyan,linkcolor=magenta}
\usepackage{multirow,array}

\usepackage{amsmath}
\usepackage{graphics}
\usepackage{epstopdf}
\usepackage{enumerate}
\usepackage{journals}

\usepackage{savesym}
\savesymbol{tablenum}
\usepackage{siunitx}
\restoresymbol{SIX}{tablenum}

\newcommand{\beq}{\begin{equation}}
\newcommand{\eeq}{\end{equation}}
\newcommand{\beqn}{\begin{eqnarray}}
\newcommand{\eeqn}{\end{eqnarray}}

\begin{document}

\title{Viscous accretion and ejection from tori around black holes in general relativity}

\author{Masaru Shibata}
\affiliation{Max-Planck-Institut f\"ur Gravitationsphysik (Albert-Einstein-Institut), Am M\"uhlenberg 1, D-14476 Potsdam-Golm, Germany}
\affiliation{Center for Gravitational Physics and Quantum Information, Yukawa Institute for Theoretical Physics, Kyoto University, Kyoto, 606-8502, Japan}

\author{Kyohei Kawaguchi}
\affiliation{Max-Planck-Institut f\"ur Gravitationsphysik (Albert-Einstein-Institut), Am M\"uhlenberg 1, D-14476 Potsdam-Golm, Germany}
\affiliation{Institute for Cosmic Ray Research, The University of Tokyo, 5-1-5 Kashiwanoha, Kashiwa, Chiba 277-8582, Japan}
\affiliation{Center for Gravitational Physics and Quantum Information, Yukawa Institute for Theoretical Physics, Kyoto University, Kyoto, 606-8502, Japan}

\author{Alan Tsz-Lok Lam}
\affiliation{Max-Planck-Institut f\"ur Gravitationsphysik (Albert-Einstein-Institut), Am M\"uhlenberg 1, D-14476 Potsdam-Golm, Germany}

\date{\today}

\begin{abstract}
We systematically perform long-term (millions of Schwarzschild time) axisymmetric viscous hydrodynamics simulations for tori around black holes in general relativity supposing the super Eddington accretion flow. The initial condition for the tori is modeled simply by the Fishbone-Moncrief torus with a constant specific angular momentum $j$ but with a wide variety of $j$.  
We find that for a given density profile, the fraction of the mass infall onto the black hole is approximately proportional to $j^{-1}$, indicating that only a minor fraction of the matter in the torus formed far from the black hole falls into the black hole while the majority is ejected with the typical average velocity of a few percent of the speed of light. We also find that the mass ejection is driven only outside $\approx 2\,r_\mathrm{ISCO}$ where $r_\mathrm{ISCO}$ is the areal radius of the innermost stable circular orbit around black holes, which depends strongly on the black hole spin. We derive an approximate fitting formula for the spin-dependence on the mass infall fraction as $\propto r_\mathrm{ISCO}^{0.7}$, which suggests that the rapid growth of supermassive black holes proceeded primarily by the accretion of the matter with the angular momentum counter-rotating with the black hole spin.
\end{abstract}

\maketitle

\section{Introduction}

The presence of supermassive black holes (SMBHs) with high estimated mass $\sim 10^7M_\odot$--$10^{10}M_\odot$ in the high-redshift universe is one of the unsolved problems in astrophysics~(see, e.g., Refs.~\cite{Fan2023aug, 2023ApJ...959...39H, Bogdan2024jan, Goulding2023sep, Kovacs2024apr}). It is natural to consider that these SMBHs were produced through a rapid mass accretion from a seed black hole of a smaller mass \cite{Inayoshi2020aug, Volonteri2021sep}. There are a wide variety of possibilities for the rapid growth of SMBHs~\cite{1984ARA&A..22..471R}. However, we still do not understand the major processes for this. In the following, we explore the growth of black holes by mass accretion from a torus surrounding the central black hole. Such a torus may be a remnant after tidal disruption of ordinary stars near relatively low-mass SMBH or an outcome of an inflow of the matter with a low specific angular momentum. 

Here, we pay attention to the viscous evolution of a torus of stellar-size mass $M_\star \sim M_\odot$ orbiting an SMBH of relatively low mass $M_\mathrm{BH}=10^5M_\odot$--$10^6M_\odot \gg M_\star$ as an example. Such a system may be formed for a high mass accretion of low-angular momentum matter or after tidal disruption of an ordinary star, which can happen near the tidal radius defined by~\cite{1975Natur.254..295H}
\beqn
r_\mathrm{t}&=& R_\star \left(\frac{M_\mathrm{BH}}{M_\star}\right)^{1/3} 
\nonumber \\
&\approx &
7.0\times 10^{12}\,\mathrm{cm}\left(\frac{R_\star}{R_\odot}\right)\left(\frac{M_\star}{M_\odot}\right)^{-1/3}
\left(\frac{M_\mathrm{BH}}{10^6 M_\odot}\right)^{1/3}\nonumber \\
&\approx & 47r_\mathrm{g}\left(\frac{R_\star}{R_\odot}\right)
\left(\frac{M_\star}{M_\odot}\right)^{-1/3}
\left(\frac{M_\mathrm{BH}}{10^6 M_\odot}\right)^{-2/3}, \label{eq2}
\eeqn
where $r_\mathrm{g}:=GM_\mathrm{BH}/c^2$ with $c$ and $G$ the speed of light and gravitational constant, respectively. As Eq.~\eqref{eq2} indicates that the radius of the tidal disruption is far from the radius of the innermost stable circular orbit $r_\mathrm{ISCO} (\leq 9 r_\mathrm{g})$ for $M_\mathrm{BH}\alt 10^6M_\odot$, and thus, we focus on the case of $r_\mathrm{t} \gg r_\mathrm{ISCO}$ in this paper. 

In reality, the torus is not formed soon after the tidal disruption since the tidal debris that eventually forms the torus is likely to have highly eccentric (nearly parabolic) orbits (see, e.g., Ref.~\cite{Rossi} for a review). The typical radius of the apocenter is~\cite{Rossi, 2013MNRAS.435.1809S}
\beqn
r_\mathrm{a}&\sim& {r_\mathrm{t}^2 \over R_\star} \nonumber \\
&\approx& 7 \times 10^{14}\,\mathrm{cm} \left({R_\star \over R_\odot}\right)
\left({M_\mathrm{BH} \over 10^6 M_\odot}\right)^{2/3}
\left({M_\star \over M_\odot}\right)^{-2/3}.~~
\eeqn
Thus, the debris should orbit the central SMBH with an orbital radius of $\alt 10^4\,r_\mathrm{g}$ interacting with each other. This suggests that the resulting torus is wide in the orbital radius with $10\,r_\mathrm{g}$--$10^4\,r_\mathrm{g}$, and thus, we suppose such a wide torus in the following. 

Assuming the shear viscosity in the form of $\nu=\alpha_\mathrm{vis} c_\mathrm{s}^2 \Omega^{-1}$ \cite{Shakura1973a}, where $\alpha_\mathrm{vis}$ is a dimensionless parameter of order 0.01, $c_\mathrm{s}$ is the sound velocity, and $\Omega$ is the angular velocity, the viscous timescale of a disk/torus is evaluated by
\beqn
t_\mathrm{vis}={R^2 \over \nu} 
&\approx& 8.9 \times 10^5\,\mathrm{s}
\left({\alpha_\mathrm{vis} \over 0.05}\right)^{-1} 
\left({M_\mathrm{BH} \over 10^6M_\odot}\right) \nonumber \\
&\times& \left({R/r_\mathrm{g} \over 10^2}\right)^{1/2}
\left({c_\mathrm{s} \over 10^9\,\mathrm{cm/s}}\right)^{-2}, \label{eq3}
\eeqn
where $R$ is the cylindrical radius and we set $\Omega=\sqrt{GM_\mathrm{BH}/R^3}$. We employ $c_\mathrm{s}=10^9\,\mathrm{cm/s}$ because we find it a typical value in the numerical computation of this paper (but near the black hole horizon, $c_\mathrm{s}$ can be $\sim 10^{10}$\,cm/s). We also note that $GM_\mathrm{BH}/c^3 \approx 4.926\,\mathrm{s}(M_\mathrm{BH}/10^6M_\odot)$, and thus, for small values of $\alpha_\mathrm{vis}\sim 0.01$, the viscous timescale exceeds  $10^6(GM_\mathrm{BH}/c^3)$, requiring us a long-term simulation. 

On the other hand, the diffusion timescale of photons may be estimated by
\beqn
t_\mathrm{diff}&=&\left({R \over \lambda}\right)^2 \left({\lambda \over c}\right)
\sim{3\kappa M_\star \over 4\pi Rc} \nonumber \\
&\approx & 2\times 10^{8}\,\mathrm{s}
\left({R \over 7 \times 10^{12}\,\mathrm{cm}}\right)^{-1}
\left({M_\star \over M_\odot}\right)
\left({\kappa \over 0.1\,\mathrm{g/cm^2}}\right), \nonumber \\
\eeqn
where $\lambda:=(\rho \kappa)^{-1}$ is the mean free path of photons with $\rho$ the rest-mass density and $\kappa$ the opacity by the Thomson scattering for an ionized fluid, and we used $\rho \sim M_\star/(4\pi R^3/3)$ for simplicity. This shows that the diffusion timescale of photons is much longer than the viscous timescale for $R \alt 10^3r_\mathrm{g}$, and hence, photons are essentially trapped, and the flow is adiabatic in the inner region.  

Furthermore, the accretion rate onto the SMBH is broadly estimated by 
\beqn
\dot M_\mathrm{BH} &\sim & {M_\star \over t_\mathrm{vis}} \nonumber \\
&\approx & 2.2 \times 10^{27}\,\mathrm{g/s} 
\left({M_\star \over M_\odot}\right)
\left({\alpha_\mathrm{vis} \over 0.05}\right) 
\left({M_\mathrm{BH} \over 10^6M_\odot}\right)^{-1} \nonumber \\
&&~~~~~~~~~~~~~~~\times \left({R/r_\mathrm{g} \over 10^2}\right)^{-1/2}
\left({c_\mathrm{s} \over 10^9\,\mathrm{cm/s}}\right)^{2},
\eeqn
which is much larger than the Eddington rate 
\beqn
\dot M_\mathrm{Edd}\approx 1.4 \times 10^{24}\,\mathrm{g/s} 
\left({M_\mathrm{BH} \over 10^6M_\odot} \right),
\eeqn
where we assumed $L_\mathrm{Edd}=\dot M_\mathrm{Edd}c^2=0.1 \dot M_\mathrm{BH} c^2$. Therefore, the accretion flow is super-Eddington. This analysis indicates that the effects of radiation transport are minor when we consider the growth of SMBHs with $M_\mathrm{BH} \alt 10^7M_\odot$ after a tidal disruption event. This is also the case for a substantial infall of mass $\sim M_\odot$ to an SMBH with a low specific angular momentum with $\alt 100 GM_\mathrm{BH}/c$.

Motivated by these facts, we study a flow in the black hole spacetime using viscous hydrodynamics without considering the radiative transport effect. Since the radiation pressure is likely to dominate over the gas pressure, we simply employ the $\Gamma$-law equation of state with $\Gamma=4/3$. Although the thermodynamics is highly simplified, we fully take into account the effects of general relativistic gravity. Furthermore, we follow the evolution of the system in the timescale of $\agt 10^6GM_\mathrm{BH}/c^3$ because Eq.~\eqref{eq3} shows that the viscous timescale is of order $10^6GM_\mathrm{BH}/c^3$ for the plausible viscous parameter and plausible remnants of tidal disruption for ordinary stars; hence, to fully understand the entire evolution of the system, i.e., the mass infall and mass ejection, such a longterm simulation is essential. 

The paper is organized as follows. In Sec.~\ref{sec2}, we describe the simulation setup in this paper. After describing the diagnostics in Sec.~\ref{sec3}, numerical results are presented in Sec.~\ref{sec4}, paying particular attention to the fraction of mass infall onto the black hole, for which we develop a simple fitting formula. We show that the general relativistic effect near the black hole is key in quantifying the mass infall fraction.  Sections~\ref{sec5} and \ref{sec6} are devoted to discussions and summary, respectively. 
In the following, unless otherwise stated, we use the units of $c=1$ for simplicity. $r_\mathrm{g}$ always denotes $GM_\mathrm{BH}/c^2$. 

\section{Simulation setup}\label{sec2}

We perform axisymmetric shear-viscous hydrodynamics simulations using a formalism shown in Ref.~\cite{Shibata:2017jyf} on a fixed background of the black hole spacetime. Following Refs.~\cite{1998ApJ...507L..67F, 1999MNRAS.305..920F, McKinney:2004ka}, the simulation is performed employing the Kerr-Schild metric. The formalism for viscous hydrodynamics is a simplified version of Ref.~\cite{Israel:1979wp}, in which the causality is guaranteed.  

In the numerical simulation, we give the kinematic shear viscous parameter $\nu$ by \cite{Shakura1973a}
\begin{equation}
\nu=\alpha_\mathrm{vis} c_\mathrm{s}^2 \Omega^{-1} f(x^i), \label{shear}
\end{equation}
where
$f(x^i)$ denotes a function of the coordinates. In this paper, we set
\begin{eqnarray}
&& \Omega^{-1}=\sqrt{{R^3 \over GM_\mathrm{BH}}},\\
&& f(x^i)={R^2 \over r^2},
\end{eqnarray}
where 
$r$ denotes the radial coordinate (in Kerr Schild coordinates). With this setting, we can give a weight for the viscosity around the equatorial region. $\alpha_\mathrm{vis}$ is chosen to be 0.01, 0.02, 0.05, and 0.1 (with 0.05 fiducial), assuming that the viscosity is caused by the magnetorotational instability and resultant dynamo in the hypothetical presence of the magnetic-field effect~\cite{Balbus:1998ja,Suzuki:2013rka,Shi:2015mvh}. We tried other choices for the function $f$ but found that the conclusion in this paper is essentially unchanged. 

We have to be careful in performing viscous hydrodynamics simulations about whether the viscous hydrodynamics is an appropriate choice, because the viscosity is believed to be effectively generated by magnetohydrodynamical turbulence in astrophysics. The turbulence in the accretion disks in magnetohydrodynamics is likely to be triggered by the magnetorotational instability~\cite{Balbus:1998ja}. The timescale for the growth of this instability until the non-linear saturation is achieved is approximately written as
\beqn
t_\mathrm{MRI}={B \over \Omega},
\eeqn
where $B$ is a dimensionless constant of order 10. This timescale has to be shorter than the viscous timescale, $t_\mathrm{vis}$ (cf.~Eq.~\eqref{eq3}). Hence, when employing the viscous hydrodynamics, the condition of $t_\mathrm{MRI} < t_\mathrm{vis}$ has to be satisfied. This is written as
\beqn
{R \over r_\mathrm{g}} < 1.8\times 10^3 
\left({\alpha_\mathrm{vis} \over 0.05}\right)^{-1}
\left({c_\mathrm{s} \over 10^9\,\mathrm{cm/s}}\right)^{-2}
\left({B \over 10}\right)^{-1}.~~~
\eeqn
We find that for the outer region $R \agt 10^3\,r_\mathrm{g}$, the sound velocity $c_\mathrm{s}$ decreases below $10^9\,\mathrm{cm/s}$, and thus the condition is a bit relaxed. However, this estimate still indicates that the viscous coefficient given by Eq.~\eqref{shear} is valid only for $R \alt 10^4\,r_\mathrm{g}$. For this reason, we set that for $R \geq 5000 \,r_\mathrm{g}$, $\nu$ in Eq.~\eqref{shear} is adjusted by changing $\Omega^{-1}$ to be a constant of $\Omega^{-1}$ at $R=5000 \,r_\mathrm{g}$. 

We employ the equation of state in the form
\beqn
P=K \rho^{4/3}+ (\Gamma-1)\rho (\varepsilon-\varepsilon_\mathrm{p}),
\eeqn
where $P$, $\rho$, $\varepsilon$, $K$, and $\Gamma$ are the pressure, rest-mass density, specific internal energy, polytropic constant, and adiabatic index, respectively. 
$\varepsilon_\mathrm{p}$ is the so-called polytropic part of the specific internal energy, written as $3K\rho^{1/3}$. Thus, for $\Gamma=4/3$, the equation of state reduces to the $\Gamma$-law equation of state, $P=(\Gamma-1)\rho \varepsilon$. We here suppose that the gas is optically thick and the pressure is determined by the radiation pressure. Throughout this paper, we basically assume that the radiation is trapped by the gas, i.e., the viscous timescale is shorter than the diffusion timescale of the radiation, because our purpose in this paper is to explore the mass accretion onto the black hole in the super Eddington regime. However, to phenomenologically investigate the effect of cooling on the efficiency of the mass infall, several simulations are performed for $\Gamma < 4/3$ (cf.~Sec.~\ref{sec4E}).

As the initial condition, we give the Fishbone-Moncrief torus~\cite{1976ApJ...207..962F}, for which the specific angular momentum and specific energy of the torus matter are constant; $\ell=-u_\varphi/u_t$ and $E=-hu_t$ are constant. Here, $h$ is the specific enthalpy, and $u_\mu$ is the lower component of the four-velocity. The specific angular momentum in the ordinary mean is defined by $j=E \ell=hu_\varphi$. For constructing the initial data, the polytropic equation of state, $P=K\rho^{4/3}$, is employed. 

It is well-known that such a torus is unstable to nonaxisymmetric deformation and subsequent angular momentum transport~\cite{1984MNRAS.208..721P, 1986PThPh..75.1464K} but we here assume that the viscous effect could phenomenologically incorporate this effect. We also note that the tori we choose have a large width so that the nonaxisymmetric instability is likely to be mild. Because $E \approx 1$ for the wide torus considered in this paper, $j \approx \ell$. Thus, in the following, each model will be specified in terms of $\hat \ell=\ell/r_\mathrm{g}$ (not $j$). 

In this paper, we choose $\hat \ell=4$, 5, 6, 8, 10, 12, and 15. The outer edge of the torus on the equatorial plane is chosen to be $\hat r_\mathrm{out}=r_\mathrm{out}/r_\mathrm{g}=10^3$ or $10^4$; we consider widely spread and fat tori. For $\hat \ell \geq 6$ the location of the inner edge, $r_\mathrm{in}$, and density maximum, $r_\mathrm{c}$, of the tori are approximately written as $\ell^2/(2r_\mathrm{g})$ and $\ell^2/r_\mathrm{g}$, respectively. Thus, for $\hat \ell \geq 10$, the density maximum is located far from the black hole, $r \agt 100\,r_\mathrm{g}$. 

We perform simulations for a wide range of the dimensionless spin parameter of black holes, $\chi=-0.8$, $-0.4$, $0$, $0.4$, $0.8$, $0.9$, and $0.95$. Here, for $\chi < 0$, the torus is counter-rotating with respect to the black hole spin. 

The simulation is performed on a two-dimensional domain of $R$ and $z$. For the $R$ and $z$ directions, a non-uniform grid is employed: For $x \alt 2.4\,r_\mathrm{g}$ ($x=R$ or $z$), a uniform grid is used, while outside this uniform region, the grid spacing $\Delta x_i$ is increased uniformly as $\Delta x_{i+1}=1.01\Delta x_i$, where the subscript $i$ denotes the $i$-th grid. Simulations are performed with the grid number of $(481, 481)$ or $(521, 521)$ for $(R, z)$; for the former and latter cases, $\Delta x$ in the innermost region is chosen as $0.06\,r_\mathrm{g}$ and $0.04\,r_\mathrm{g}$, respectively. 
The black-hole horizon, for which the coordinate radius is written as $r_\mathrm{H}=r_\mathrm{g}(1+ \sqrt{1-\chi^2})$, is always located in the uniform grid zone, and the outer boundaries along the $R$ and $z$ axes are located at $\approx 1.85 \times 10^4\,r_\mathrm{g}$. We confirm that the mass of the matter swallowed into the black hole depends only weakly on the grid resolution unless the value of $\alpha_\mathrm{vis}$ is as large as $0.1$. 

In this paper, we employ the code developed in Ref.~\cite{Shibata:2017jyf}. Some of the simulations were also performed using a new fixed-mesh refinement code developed independently by one of the authors (Lam et al., in preparation; see also Ref.~\cite{Lam:2025pmz}). We confirmed that the results from the two independent codes agree well. 

\section{Diagnostics}\label{sec3}

During each run, we monitor the rates of the rest mass that falls into the black hole and is ejected from the system respectively, by
\beqn
\dot M_\mathrm{BH}&=& -\oint_{r=r_\mathrm{H}} \rho \sqrt{-g} u^r dS,\\
\dot M_\mathrm{eje}&=& \oint_{r=r_\mathrm{o}} \rho \sqrt{-g} u^r dS,
\eeqn
where $g$ denotes the determinant of the spacetime metric, $dS=d\theta d\varphi$, and we choose the extraction radius of the ejected matter as $r_\mathrm{o}=5000\,r_\mathrm{g}$. We integrate these quantities in time to get the total mass swallowed by the black hole and that of the ejecta as
\beqn
\Delta M_* &=& \int dt \dot M_\mathrm{BH},\\
M_\mathrm{eje} &=& \int dt \dot M_\mathrm{eje}. 
\eeqn
In the same way, the quantities associated with the angular momentum and energy can be calculated by replacing $\rho \sqrt{-g} u^r$ to $\rho h u_\varphi \sqrt{-g}u^r$ and $-\rho h u_t \sqrt{-g}u^r$, respectively. We denote the infall rates of the angular momentum and energy into the black hole by $\dot J_\mathrm{BH}$ and $\dot E_\mathrm{BH}$, respectively, and the ejection rate of the energy by $\dot E_\mathrm{eje}$. From $\dot E_\mathrm{eje}$ and $\dot M_\mathrm{eje}$, we define the average velocity of the ejecta as
\beqn
v_\mathrm{eje}=\sqrt{2\left(\left\langle{\dot E_\mathrm{eje} \over \dot M_\mathrm{eje}}\right\rangle-1-W_\mathrm{o}\right)},
\eeqn
where $W_\mathrm{o}$ denotes the specific gravitational potential energy for which we simply set $W_\mathrm{o}=GM_\mathrm{BH}/r_\mathrm{o}$. $\langle\cdots\rangle$ denotes the time averaging, which is taken because $\dot E_\mathrm{eje}/\dot M_\mathrm{eje}$ varies in a short timescale. 
In the above definition, we supposed that the internal energy of the ejecta is much smaller than the kinetic energy at large radii. In our present setup, we find $\dot E_\mathrm{eje}/\dot M_\mathrm{eje} > 1 + W_\mathrm{o}$ for most cases, and hence, the ejecta velocity is measured. However, for $\alpha_\mathrm{vis}=0.01$ or $\chi < 0$, $\dot E_\mathrm{eje}/\dot M_\mathrm{eje}$ can be close to or less than $1 + W_\mathrm{o}$ in a late stage of the evolution, although for most stages, we can still measure the ejecta velocity (cf. Sec.~\ref{sec4C}). 

We analyze the mass of the inflowing matter at each radius $r$ by defining 
\beqn
\dot M_*(r)&=& -\oint_{r} \rho \sqrt{-g} u^r dS, \\
\Delta M_*(r)&=& \int dt \dot M_*(r).
\eeqn
These quantities are considered as the {\it net} inflow mass (if $\dot M_*(r)$ is positive). We also define the {\it gross} mass infall rate of the inflowed matter, i.e., only with $u^r <0$, by
\beqn
\dot M_*^\mathrm{in}(r)&=& -\oint_{r, u^r<0} \rho \sqrt{-g} u^r dS. 
\eeqn
From $\dot M_*(r)$ and $\dot M_*^\mathrm{in}(r)$ we can evaluate the outflow rate at $r$ by $\dot M_*^\mathrm{out}(r)=\dot M_*^\mathrm{in}(r)-\dot M_*(r)$. 
In the following, we pay particular attention to the normalized quantities such as $\Delta M_*/M_*$ and $\Delta M_*(r)/M_*$ where $M_*$ denotes the initial total baryon mass. 

Using the quantities at the black hole horizon, we can determine the evolution of the dimensionless spin of the black hole by~\cite{2004ApJ...602..312G}
\beqn
\Delta \chi &=&{GM_\mathrm{BH}^2 \chi + \Delta J_\mathrm{BH} \over G(M_\mathrm{BH} + \Delta E_\mathrm{BH})^2}-\chi \nonumber \\
&\approx& {\Delta J_\mathrm{BH} -2 \chi \Delta E_\mathrm{BH} G M_\mathrm{BH} \over GM_\mathrm{BH}^2}, \label{deltachi}
\eeqn
where $\Delta J_\mathrm{BH}$ and $\Delta E_\mathrm{BH}$ are obtained by the time integral of $\dot J_\mathrm{BH}$ and $\dot E_\mathrm{BH}$, respectively, and 
we supposed that $\Delta E_\mathrm{BH} \ll M_\mathrm{BH}$. Equation~\eqref{deltachi} is written in the form of the time evolution of the dimensionless spin $\dot\chi$ as
\beqn
\dot \chi \approx {\dot J_\mathrm{BH} -2 \chi \dot E_\mathrm{BH} G M_\mathrm{BH} \over G M_\mathrm{BH}^2}, \label{deltachi2}
\eeqn
and thus, we monitor a dimensionless quantity defined by 
\beqn
\zeta
:={\dot J_\mathrm{BH} -2 \chi \dot E_\mathrm{BH} G M_\mathrm{BH} \over G M_\mathrm{BH} \dot M_\mathrm{BH}}, \label{deltachi3}
\eeqn
to analyze whether the black hole spins up ($\zeta > 0$) or down ($\zeta < 0$). 

We note that if the matter falls into a black hole adiabatically from the innermost stable circular orbit, $\zeta$ is written as $\zeta=\zeta_\mathrm{ISCO}=\ell_\mathrm{ISCO}/r_\mathrm{g}-2\chi e_\mathrm{ISCO}$ where $\ell_\mathrm{ISCO}(\geq 2\,r_\mathrm{g}/\sqrt{3})$ and $e_\mathrm{ISCO}(\geq 1/\sqrt{3})$ are the specific angular momentum and specific energy, respectively, of a test particle at the innermost stable circular orbits. $\zeta_\mathrm{ISCO}$ depends on $\chi$~\cite{1972ApJ...178..347B, 1983bhwd.book.....S} but it is always positive for $-1 \leq \chi < 1$ ($\zeta_\mathrm{BH}=0$ for $\chi=1$). 

%%%%%%%%%%%%
\section{Results} \label{sec4}

\subsection{General feature}\label{sec4.1}

\begin{figure}
\includegraphics[width=0.495\textwidth]{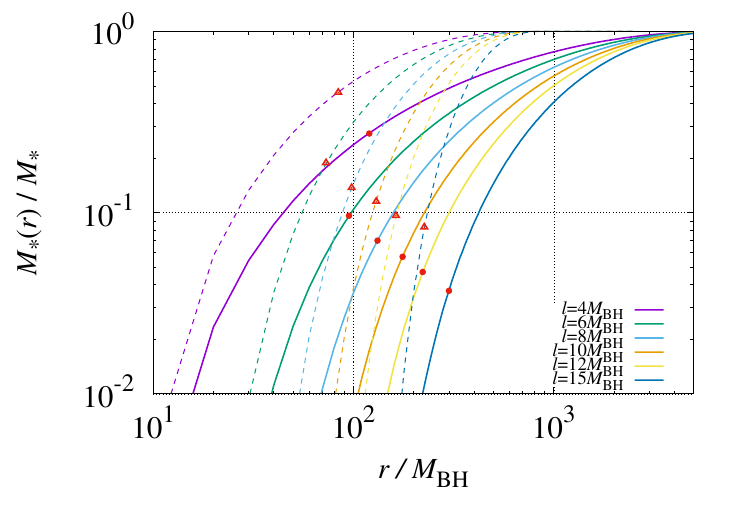}
\vspace{-10mm}
\caption{The profile of $M_*(r)$ in units of the entire mass $M_{*}$ as a function of $r/r_\mathrm{g}$ for a variety of $\hat\ell$ with $\chi=0$ and $\hat r_\mathrm{out}=10^4$ (solid curves) and $10^3$ (dashed curves). We note that the curves are not modified significantly even if we change the dimensionless spin. The circles and triangles show the numerical results for the mass fraction of the matter that is swallowed into the black hole, obtained in this paper for $\alpha_\mathrm{vis}=0.05$, and are located at $r_\mathrm{c}$--$2r_\mathrm{c}$ for $\hat\ell \agt 6$. Here, $r_\mathrm{c}$ denotes the radius of the density maximum on the equatorial plane.
}
\label{fig1}
\end{figure}

\begin{figure*}
\hspace{-5mm}
\includegraphics[width=0.51\textwidth]{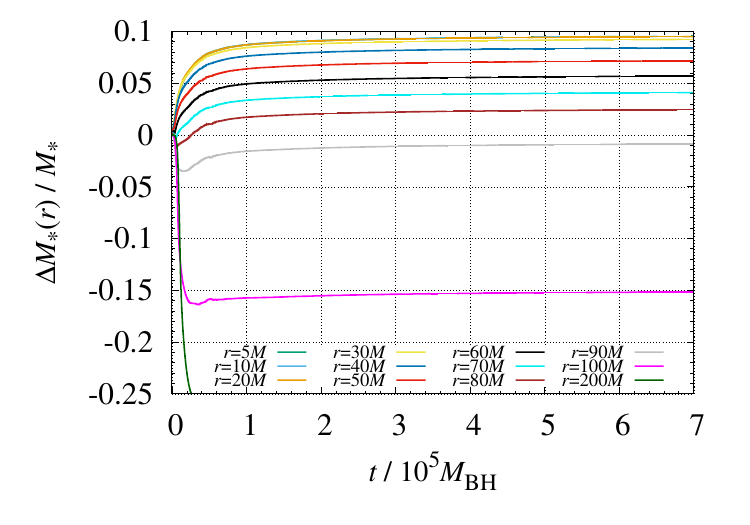}
\hspace{-5mm}
\includegraphics[width=0.51\textwidth]{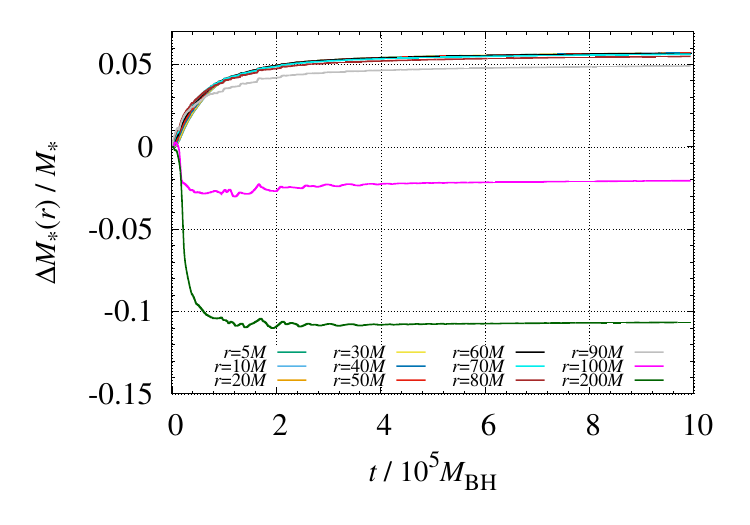}
\vspace{-4mm}
\caption{The evolution of the mass that falls inside a radius of $r$, $\Delta M_*(r)$, as a function of time, $t/(10^5GM_\mathrm{BH})$, for a variety of $r$ for the models with $\chi=0$,  $\hat r_\mathrm{out}=10^4$, $\alpha_\mathrm{vis}=0.05$, and $\hat \ell=6$ (left) and $\hat \ell=10$ (right).}
\label{fig2}
\end{figure*}

\begin{figure*}
\includegraphics[width=0.49\textwidth]{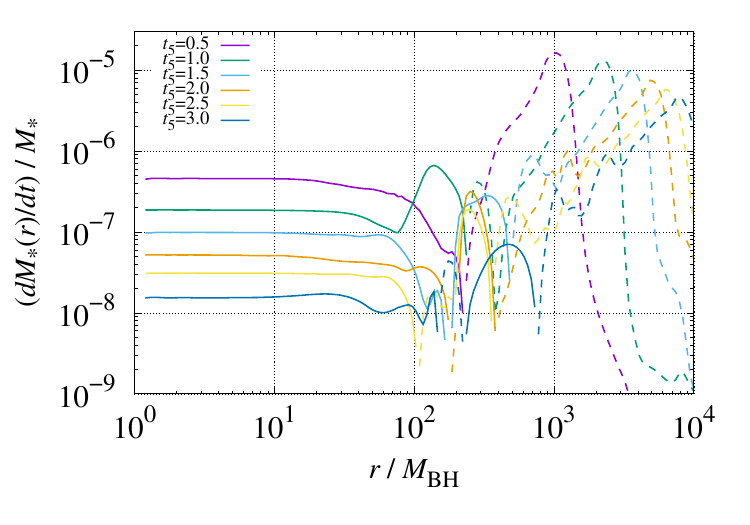}
\includegraphics[width=0.49\textwidth]{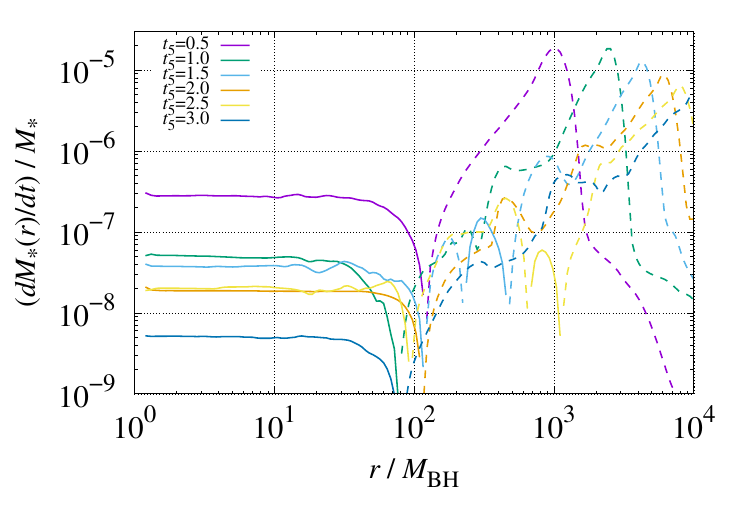}\\
\vspace{-6mm}
\includegraphics[width=0.49\textwidth]{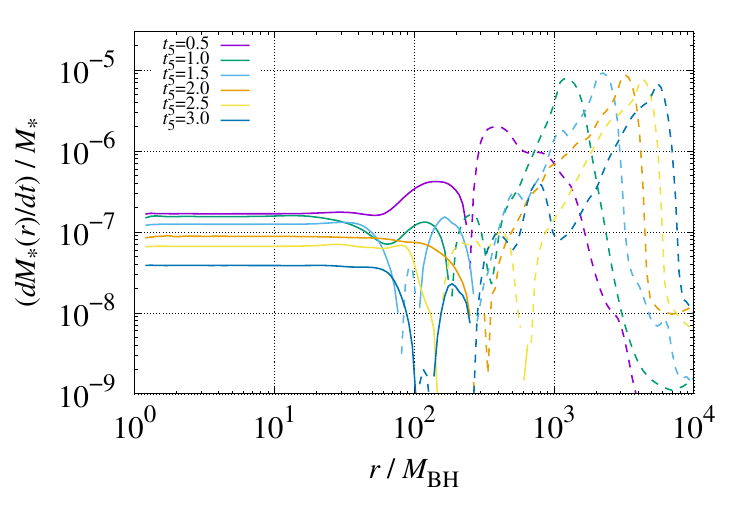}
\includegraphics[width=0.49\textwidth]{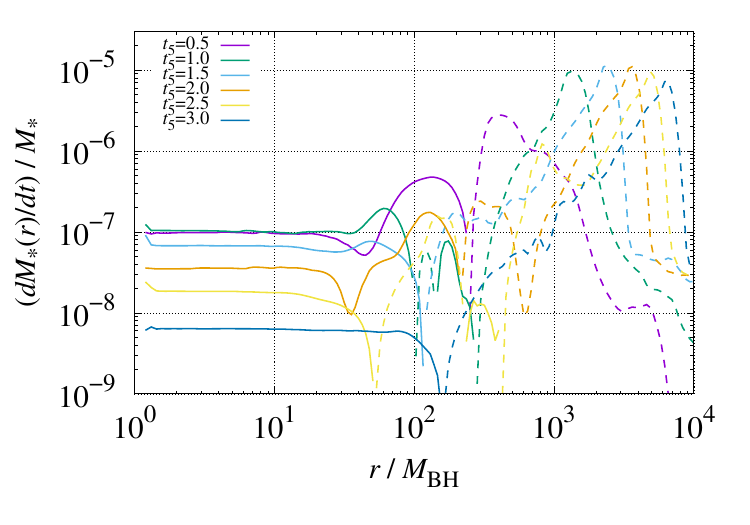}\\
\vspace{-6mm}
\includegraphics[width=0.49\textwidth]{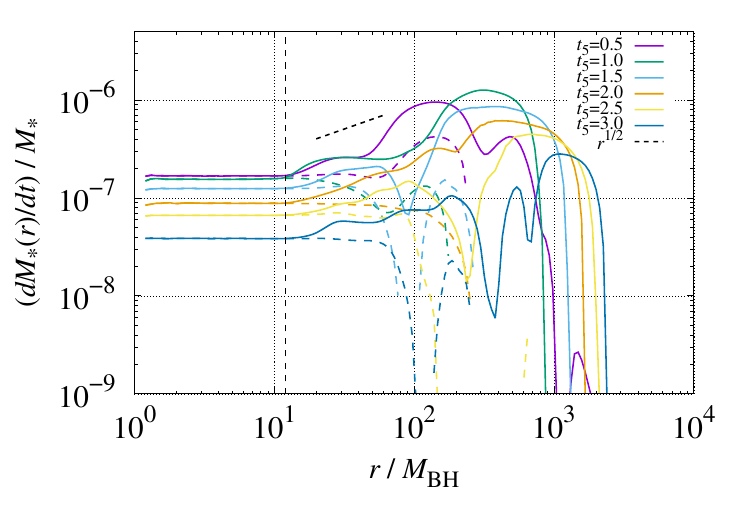}
\includegraphics[width=0.49\textwidth]{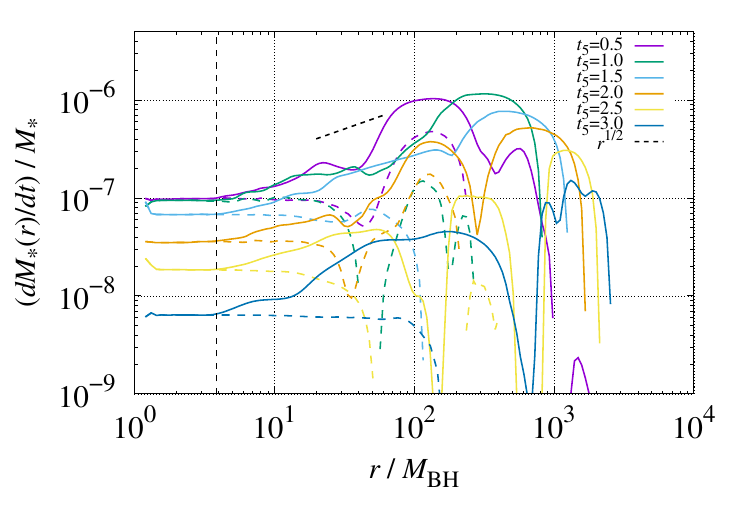}
\vspace{-4mm}
\caption{Mass infall (outflow) rates $\dot M_*(r)$ as functions of $r/r_\mathrm{g}$ at selected time slices for $(\hat\ell, \chi)=(10, 0)$ (top left), $(10, 0.95)$ (top right), $(15, 0)$ (moddel left), and $(15, 0.95)$ (middle right). The solid and dashed curves are plotted when $\dot M_*(r)$ is positive and negative, respectively. The bottom two panels show the comparison of the net mass infall rate, $\dot M_*(r)$ (dashed curve), and gross one, $\dot M_*^\mathrm{in}(r)$ (solid curve), for $(\hat\ell, \chi)=(15, 0)$ (left) and $(15, 0.95)$ (right). For all the cases, $\hat r_\mathrm{out}=10^4$ and $\alpha_\mathrm{vis}=0.05$. $t_5$ denotes $t/(10^5GM_\mathrm{BH})$. The vertical dashed lines and dashed slopes in the bottom two panels denote $r=2\,r_\mathrm{ISCO}$ and $\propto r^{1/2}$, respectively. 
}
\label{fig25}
\end{figure*}

\begin{figure*}
\hspace{-8mm}
\includegraphics[width=0.52\textwidth]{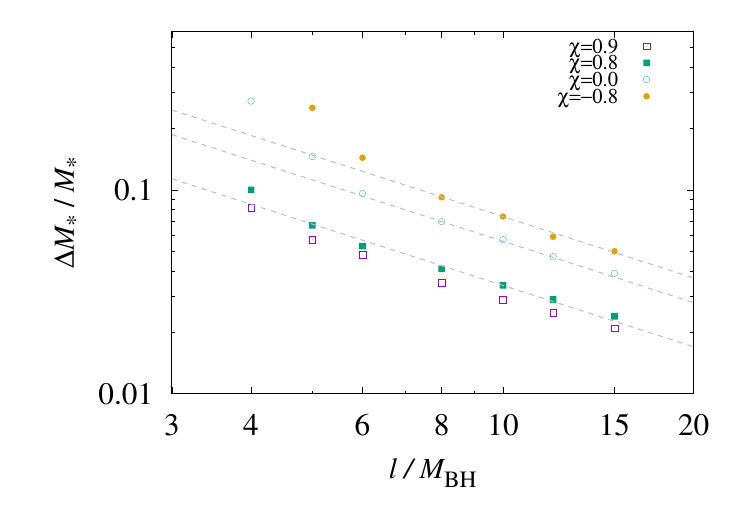}
\hspace{-8mm}
\includegraphics[width=0.52\textwidth]{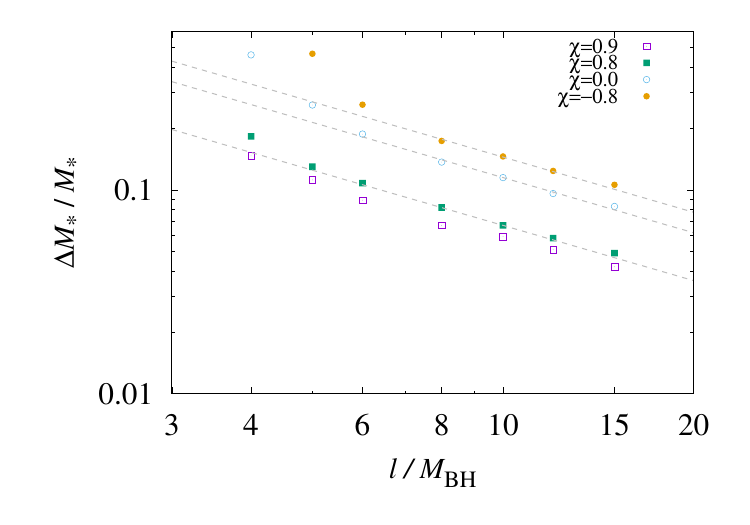}
\vspace{-5mm}
\caption{The total mass fraction of the matter swallowed into the black hole as a function of $\hat\ell$ for $\chi=0.9$, 0.8, 0, and $-0.8$ with $\alpha_\mathrm{vis}=0.05$ and $\hat r_\mathrm{out}=10^4$ (left) and $10^3$ (right). The dashed slopes are $\propto \ell^{-1}$ (left) and $\ell^{-0.9}$ (right).
}
\label{fig3}
\end{figure*}

Irrespective of the torus configuration and black hole spin, the torus universally evolves by the viscous effect broadly in the following manner if $\ell$ is not extremely small as $\ell \alt 5 r_\mathrm{g}$: A substantial fraction of the matter initial located for $R \alt r_\mathrm{c}\approx \ell^2/r_\mathrm{g}$ falls into the black hole while a fraction of the matter in the outer region of $R > r_\mathrm{c}$ falls toward the black hole and another substantial fraction simply spreads outward, eventually becoming ejecta. Here, the mass ejection is driven by the viscous heating of the matter located in the vicinity of the black hole for which the viscous heating is most efficient, as we found in our previous work~\cite{Fujibayashi2020a}: The generated heat near the black hole produces a hot bubble at the inner region of the torus. Subsequently, the hot bubble is moved outward as the convective motion pushes the matter in the outer part of the torus with $R > r_\mathrm{c}$ outward. This pushing induces the expansion of the torus, and the matter in the outer part of the torus eventually becomes the ejecta. The mass ejection timescale is approximately described by Eq.~\eqref{eq3}, which may be written as
\beqn
{t_\mathrm{vis} \over GM_\mathrm{BH}} &\approx& 1.8\times 10^5 
\left({\alpha_\mathrm{vis} \over 0.05}\right)^{-1} \nonumber \\
&&\times \left({r_\mathrm{c}/r_\mathrm{g} \over 10^2}\right)^{1/2}
\left({c_\mathrm{s} \over 10^9\,\mathrm{cm/s}}\right)^{-2}. \label{eq31}
\eeqn

We note that the mass ejection is driven by viscous heating, which is strongest in the vicinity of the black hole. This implies that accurately resolving the viscous matter motion near the black hole is key to obtaining a reliable result in this problem. We find that our results on $\Delta M_*$ depend only weakly on the grid resolution which we employ, in particular for $\alpha_\mathrm{vis} \leq 0.05$. 

Figure~\ref{fig1} shows the initial mass distribution of the torus for $\chi=0$ as a function of $r$ defined by 
\beqn
M_*(r)=\int_{r'<r} \rho \sqrt{-g} u^t d^3x'. 
\eeqn
Each curve shows the results for the different values of $\hat \ell(=4, 6, 8, 10, 12, 15)$, and the solid and dashed curves denote the cases of $\hat r_\mathrm{out}=10^4$ and $10^3$, respectively. The circles and triangles show the total mass swallowed into the black hole, $\Delta M_*$, obtained by the numerical simulations with $\alpha_\mathrm{vis}=0.05$. 
This figure shows that the circles and triangles are located at $r=r_\mathrm{c}$--$2r_\mathrm{c}$ for $\hat \ell \geq 6$, indicating that only the matter located close to $r_\mathrm{c}$ is swallowed into the black hole. On the other hand, for the small values of $\hat\ell$ such as $\hat \ell=4$, a substantial amount of the matter with $\agt 0.2M_{*}$ falls into the black hole even from a far outside of the radius of $r_\mathrm{c}$. This is reasonable because $\hat \ell=4$ is close to the value at the innermost stable circular orbit ($\hat \ell_\mathrm{ICSO}=\ell_\mathrm{ISCO}/r_\mathrm{g}$ which is $2\sqrt{3}$ for $\chi=0$), and hence, a small fraction of the angular momentum transport for the matter can induce the matter infall into the black hole.

Figure~\ref{fig2} plots the evolution of the mass that falls inside a radius of $r$, $\Delta M_*(r)$, as a function of time ($t_5:=t/(10^5GM_\mathrm{BH}$) for a variety of $r$ for the models with $\chi=0$, $\hat r_\mathrm{out}=10^4$, $\alpha_\mathrm{vis}=0.05$, and $\hat \ell=6$ (left) and $\hat \ell=10$ (right). We note that $\Delta M_*(r)$ is obtained by the sum of both the inflow and outflow components. Both the left and right panels show that for $r \alt r_\mathrm{c}$, which is, respectively, $\sim 30\,r_\mathrm{g}$ (left) and $\sim 100\,r_\mathrm{g}$ (right), the values of $\Delta M_*(r)$ are approximately identical for each case. This appears to indicate that once the matter falls inside $r_\mathrm{c}$, they might eventually fall into the black hole, but actually, this interpretation is not correct (see below). Figure~\ref{fig2} also shows that for large values of $r$ with $r > r_\mathrm{c}$, $\Delta M_*(r)$ is negative, implying that most of the matter located initially in the outer region is ejected from the system (this is indeed the case; see below). These features are universally seen irrespective of the values of $\chi$, $\hat\ell$, $\hat r_\mathrm{out}$, and $\alpha_\mathrm{vis}$. For $\hat \ell \agt 10$, we always find that $\Delta M_*(r) < 0$ for $r \agt r_\mathrm{c}$; the matter located in such an outer region is simply ejected from the system. 

The top and middle rows of Fig.~\ref{fig25} show snapshots of $\dot M_*(r)$ at selected time slices for $(\hat\ell, \chi)=(10, 0)$ (top left), $(10, 0.95)$ (top right), $(15, 0)$ (middle left), and $(15, 0.95)$ (middle right). For all the cases, $\hat r_\mathrm{out}=10^4$ and $\alpha_\mathrm{vis}=0.05$. We note again that $\dot M_*(r)$ is obtained by the sum of both the inflow and outflow components. The solid and dashed curves are plotted when $\dot M_*(r)$ is positive and negative, respectively. Since $\dot M_*(r)$ varies with time, we take the simple average for the time interval of $[t-\Delta t,t]$ where $\Delta t \approx 3000GM_\mathrm{BH}$. It is found that irrespective of $\ell$ and $\chi$, $\dot M_*(r)$ is approximately constant at given time in an inner region of $\alt 30$--$80\,r_\mathrm{g}$~\cite{2014MNRAS.439..503S, 2014ApJ...796..106J, 2015MNRAS.447...49S, 2016MNRAS.456.3929S, 2018PASJ...70..108K, 2022PASJ...74.1378Y, 2022ApJ...934..132H}, although the mass accretion rate decreases with time in our present setting. Again, this appears to indicate that the mass ejection from the inner region might be minor, but this is not the case (see below). $\dot M_*(r)$ is smaller for larger values of $\chi$ at given time, reflecting the fact that the mass accretion is suppressed by the corotating spin effect because the innermost stable orbit is closer to the black hole for the larger values of $\chi$ (cf. Sec.~\ref{sec4b}). 

For the majority of the outer region with $r > r_\mathrm{c} \approx 100\,r_\mathrm{g}$--$200\,r_\mathrm{g}$, $\dot M_*(r)$ is always negative, showing that a substantial fraction of the matter initially located in such outer regions is ejected from the system. The peak value of $\dot M_*(r)$ for the ejecta component decreases with the ejecta going outward. This implies that a faction of the matter falls back toward the central region. However, the majority of such matter is ejected from the central region again by the viscous heating. These ejecta components are seen as the second and third peaks of the dashed curves at later stages of the evolution in Fig.~\ref{fig25}. 

To clarify how the matter is actually outflowed from the torus, we compare $\dot M_*(r)$ (net inflow rate) and $\dot M_*^\mathrm{in}(r)$ (gross inflow rate) for $(\hat\ell, \chi)=(15, 0)$ (left) and $(15, 0.95)$ (right) in the bottom two panels of Fig.~\ref{fig25}. The difference, $\dot M_*^\mathrm{in}(r)-\dot M_*(r)$, is considered as the outflow rate. It is found that $\dot M_*(r)$ is approximately equal to $\dot M_*^\mathrm{in}(r)$ only in an innermost region of $r \leq r_\mathrm{cap}$ where $r_\mathrm{cap} \approx 2\,r_\mathrm{ICSO}=12\,r_\mathrm{g}$ and $\approx 3.9\,r_\mathrm{g}$ for $\chi=0$ and $0.95$, respectively. Here $\hat r_\mathrm{ISCO}:=r_\mathrm{ISCO}/r_\mathrm{g}$ depends strongly on $\chi$ (see also Figs.~\ref{fig7} and \ref{fig8} for $\chi=\pm 0.8$). This shows that only the matter captured in the innermost region of $\alt r_\mathrm{cap}$ is swallowed by the black hole without outflow. This implies that a substantial fraction of the matter moves inward inside $r=r_\mathrm{c}$, but the majority is ejected from the system for $r \agt r_\mathrm{cap}$. This result is essentially the same as in a Newtonian simulation~\cite{1999MNRAS.310.1002S}. $r_\mathrm{cap}$ does not depend on the equations of state and $\alpha_\mathrm{vis}$; i.e., it is determined purely by the general relativistic gravity (see subsequent subsections).

Outside the capture radii $r_\mathrm{cap}$, $M_*^\mathrm{in}(r)$ is a slowly increasing function of $r$; compare with the dashed slope of $\propto r^{1/2}$ indicating that the flow is convective dominant~\cite{1999MNRAS.310.1002S,  2000ApJ...539..798N, 2000ApJ...539..809Q, 2014ARA&A..52..529Y, 2022ApJ...934..132H, 2024MNRAS.532.4826T}. We will discuss the dependence of $M_*^\mathrm{in}$ on $r$ in more detail in Sec.~\ref{sec4b}. For $r \agt 2\,r_\mathrm{cap}$, $\dot M_*(r)$ is much smaller than $\dot M_*^\mathrm{in}(r)$, and thus, the inflowed rate is comparable to the outflowed rate; i.e., most of the inflowed matter is eventually ejected for $r \agt2\,r_\mathrm{cap}$. For $r \agt r_\mathrm{c}$, $\dot M_*(r) \leq 0$. This implies that a majority of the matter located in the outer region is eventually ejected from the system. 

The captured region is smaller for the larger values of $\chi$ (see Sec.~\ref{sec4b}), for which the radius of the innermost stable circular orbit is smaller and the viscous heating rate is higher. As a result of these effects, the mass ejection rate becomes higher.

In an intermediate region around $r \sim r_\mathrm{c}$, the variability of the mass accretion (and ejection) rates is high. This reflects the fact that for such a region both mass inflow and ejection occur in an irregular manner. In the late stage of the evolution, the mass inflow is caused by the fall-back of the matter.

The local viscous heating rate is approximately proportional to $GM_\mathrm{BH} \dot M_*^\mathrm{in}(r)/r$. Since $\dot M_*^\mathrm{in}(r)$ is approximately constant for $r \alt r_\mathrm{cap}$, the viscous heating rate is proportional to $r^{-1}$ and is higher for smaller values of $r$. This is the reason that the convective motion is induced from the innermost region of the torus. It is also worth noting that it is essential to resolve the matter motion and viscous heating in the vicinity of the black hole in this problem. Newtonian and pseudo-Newtonian simulations often excise the inner region far from the black hole radius and impose an artificial boundary condition on the excised radius (e.g., Refs.~\cite{2022ApJ...934..132H, 2024MNRAS.532.4826T}). In such simulations, the viscous heating is likely to be underestimated, and thus, the mass accretion fraction may be overestimated.

%% spin dependence
\subsection{Dependence of $\Delta M_*$ on $\ell$, $\chi$, and $\alpha_\mathrm{vis}$} \label{sec4b}

\begin{figure}
\includegraphics[width=0.5\textwidth]{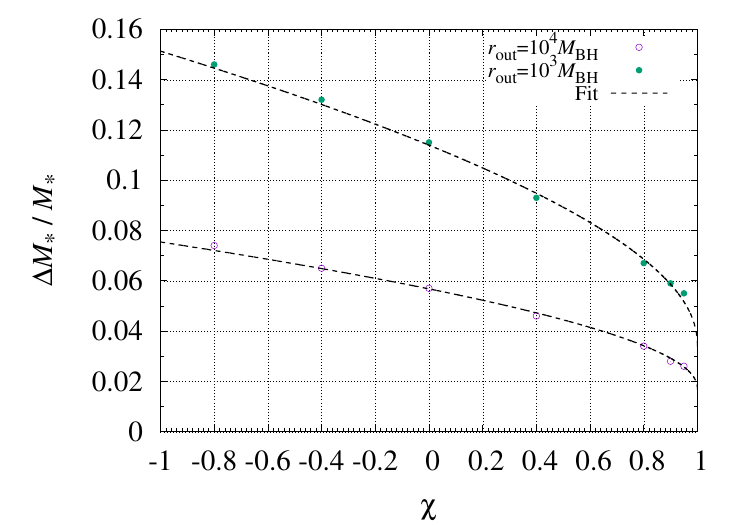}
\vspace{-5mm}
\caption{$\Delta M_*/M_*$ for $\hat\ell=10$ with $\hat r_\mathrm{out}=10^4$ (lower sequence) and $10^3$ (upper sequence) as a function of $\chi$. $\alpha_\mathrm{vis}=0.05$. 
The upper and lower dashed curves are $0.0325\,\hat r_\mathrm{ISCO}^{0.7}$ and $0.0162\,\hat r_\mathrm{ISCO}^{0.7}$, respectively.}
\label{fig4}
\end{figure}

\begin{figure*}
\hspace{-4mm}
\includegraphics[width=0.5\textwidth]{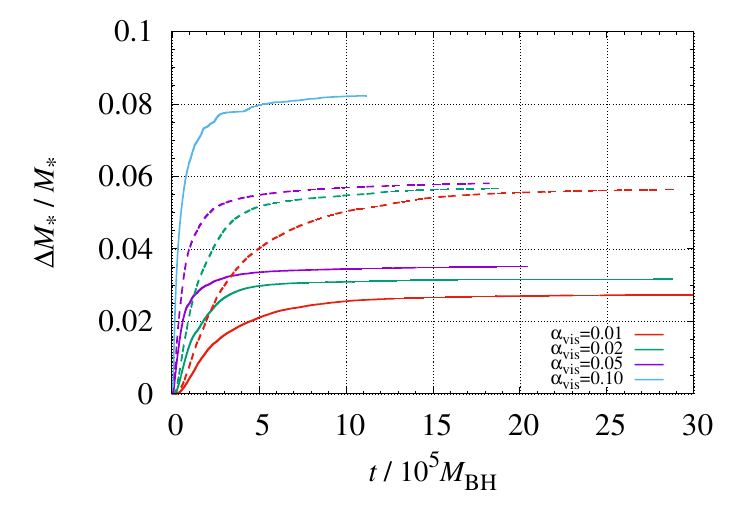}
\hspace{-5mm}
\includegraphics[width=0.5\textwidth]{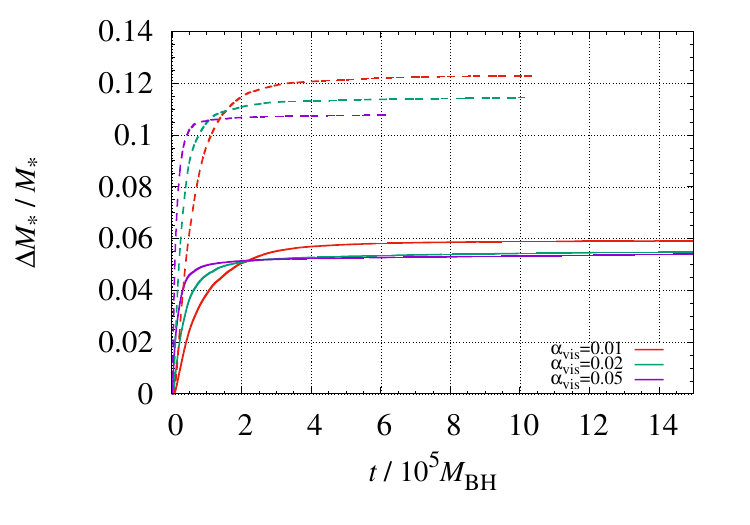}\\
\vspace{-4mm}
\includegraphics[width=0.5\textwidth]{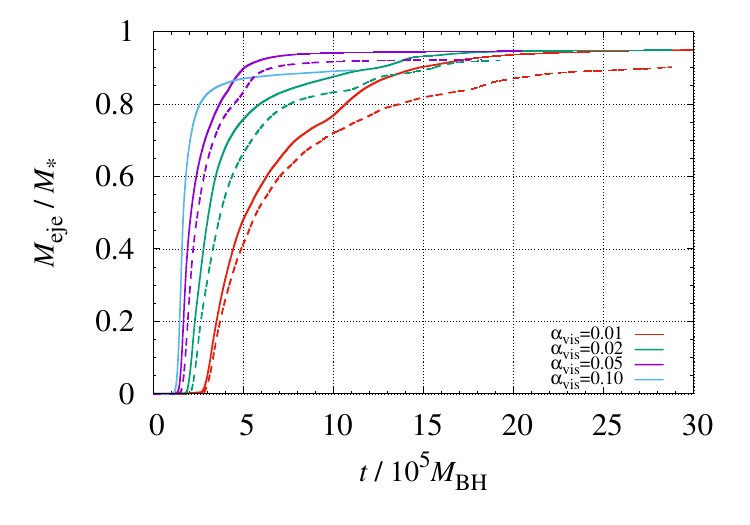}
\hspace{-5mm}
\includegraphics[width=0.5\textwidth]{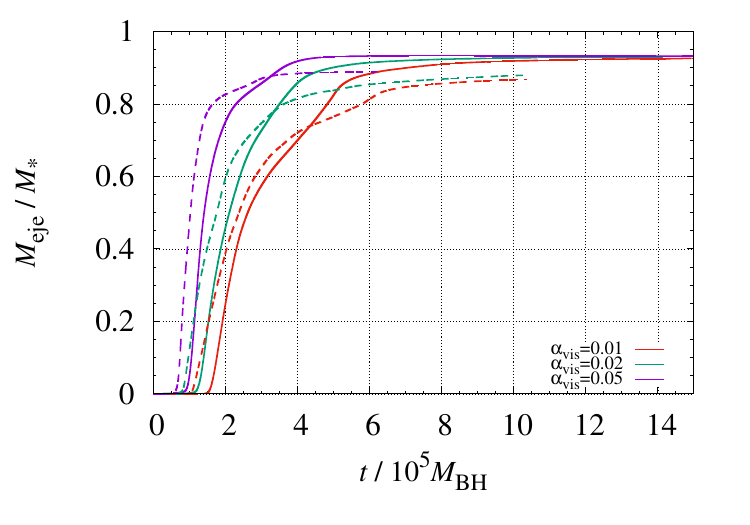}
\vspace{-4mm}
\caption{$\Delta M_*/M_*$ (upper) and $M_\mathrm{eje}/M_*$ (lower) as functions of time, $t/(10^5GM_\mathrm{BH})$, for $\hat \ell=10$ and $\hat r_\mathrm{out}=10^4$ with $\alpha_\mathrm{vis}=0.01$, 0.02, 0.05, 0.10, and $\chi=0.8$ (left, solid), with $\alpha_\mathrm{vis}=0.01$, 0.02, 0.05, and $\chi=0$ (left, dashed), and for $\hat \ell=6$ and $\chi=0.8$ with $\alpha_\mathrm{vis}=0.01$, 0.02, 0.05, and $\hat r_\mathrm{out}=10^4$ (right, solid) and $10^3$ (right, dashed).
}
\label{fig5}
\end{figure*}

\begin{figure}
\includegraphics[width=0.46\textwidth]{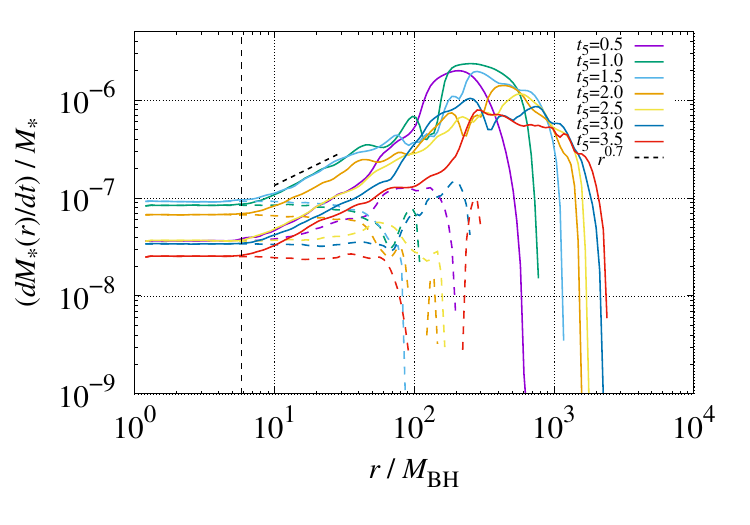}\\
\vspace{-5mm}
\includegraphics[width=0.46\textwidth]{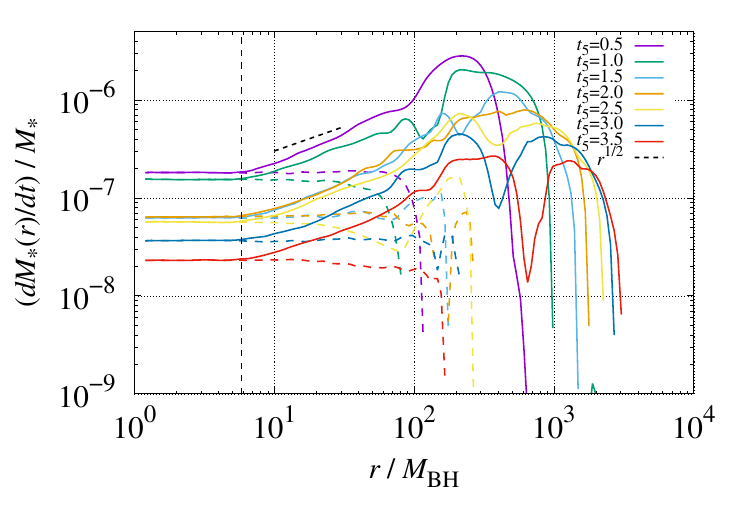}\\
\vspace{-5mm}
\includegraphics[width=0.46\textwidth]{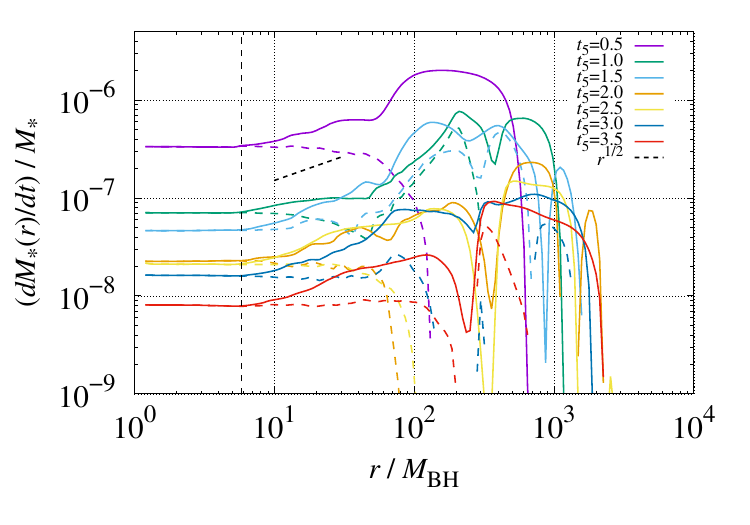}
\vspace{-5mm}
\caption{The same as the bottom panels of Fig.~\ref{fig25} but for $\hat \ell=10$, $\chi=0.8$, $\hat r_\mathrm{out}=10^4$, and $\alpha_\mathrm{vis}=0.01$ (top), 0.02 (middle), and 0.05 (bottom). The dashed slope is $\propto r^{0.7}$ for the top panel and $r^{1/2}$ for the middle and bottom panels. The vertical dashed lines show $2\,r_\mathrm{ISCO}\approx 5.8\,r_\mathrm{g}$. We note that for $\alpha_\mathrm{vis}=0.01$ the evolution is still in the middle of a relaxation at $t_5=0.5$.}
\label{fig7}
\end{figure}

\begin{figure}
\includegraphics[width=0.46\textwidth]{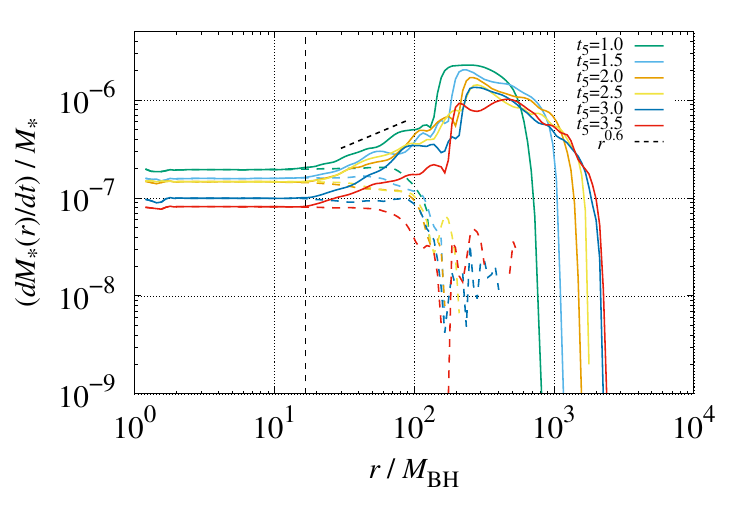}\\
\vspace{-5mm}
\includegraphics[width=0.46\textwidth]{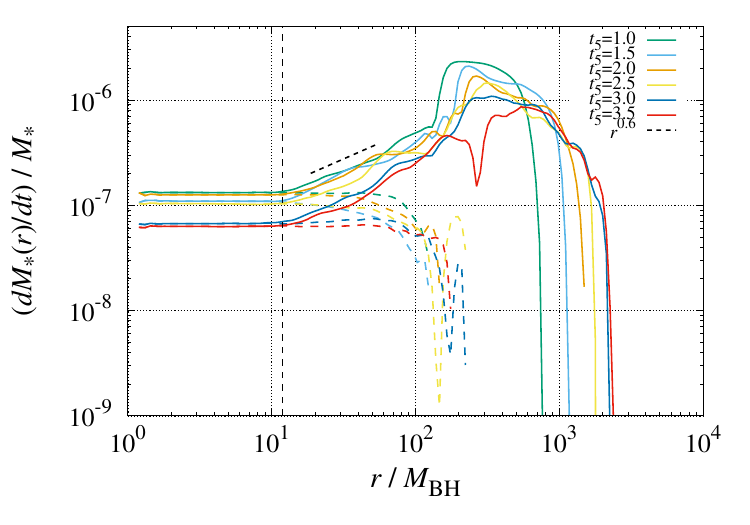}\\
\vspace{-5mm}
\includegraphics[width=0.46\textwidth]{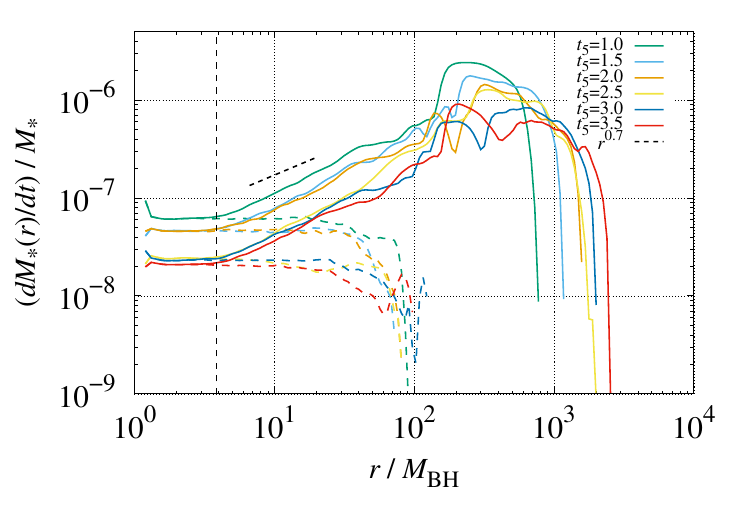}
\vspace{-5mm}
\caption{The same as the top panels of Fig.~\ref{fig7} (i.e., $\hat \ell=10$, $\hat r_\mathrm{out}=10^4$, and $\alpha_\mathrm{vis}=0.01$) but for $\chi=-0.8$ (top), $0$ (middle), and $0.95$ (bottom) with $t_5 \geq 1$. 
For the top, middle, and bottom panels, the dashed slope is $0.6$, 0.6, and 0.7, respectively.}
\label{fig8}
\end{figure}

Figure~\ref{fig3} plots the total mass fraction of the matter swallowed into the black hole, $\Delta M_*/M_*$, as a function of $\hat\ell$ with $\alpha_\mathrm{vis}=0.05$. We find that the mass fraction depends appreciably on the dimensionless spin of the black hole; it increases by a factor of several if we change it from $\chi=0.9$ to $-0.8$. This factor is, in particular, large for small values of $\hat\ell$. We find that the corotating spin significantly suppresses the matter infall onto the black hole. 

We also find that the mass infall fraction decreases systematically with the increase of $\hat\ell$. For $\hat r_\mathrm{out}=10^4$ and $10^3$ the dependence is approximately described by $\ell^{-1}$ and $\ell^{-0.9}$, respectively. This shows that for a swarm of matter with larger specific angular momentum, the mass infall into the black hole is less subject. An implication of this result will be discussed in Sec.~\ref{sec5}. 

Our results for the mass fraction that falls into the black hole, $O(0.1)$, agree broadly with the results by radiation magnetohydrodynamics simulations in general relativity~\cite{2015MNRAS.447...49S, 2016MNRAS.456.3929S}, which also find that only a small fraction of the torus matter falls into the black hole while a substantial fraction is ejected from the system. Our results are also consistent with viscous radiation hydrodynamics results for the high mass accretion cases~\cite{2018PASJ...70..108K}, which shows that the mass infall fraction decreases below 0.1 with the increase of the mass injection rate, i.e., with the decrease of the radiation cooling efficiency.  

To quantify a strong dependence of the mass fraction of the matter swallowed into the black on the dimensionless spin $\chi$, we plot $\Delta M_*/M_*$ for $\hat\ell=10$ with $\hat r_\mathrm{out}=10^3$ and $10^4$ as functions of $\chi$ as an example in Fig.~\ref{fig4}. 
This shows that $\Delta M_*/M_*$ indeed depends significantly on the dimensionless spin; high dimensionless spins corotating with the torus suppress the fraction of the matter that falls into the black hole. This suggests that the growth of the black hole would be achieved predominantly by the infall of the matter for which the direction of the angular momentum is opposite to the black hole spin. 

Together with the numerical data, we plot the curves which approximately reproduce the numerical data in Fig.~\ref{fig4}. These curves are derived based on the numerical results of Fig.~\ref{fig25} (see also the discussion below), which indicate that $\dot M_*(r)$ at the event horizon is determined by $\dot M_*^\mathrm{in}$ at $r=r_\mathrm{cap} \approx 2\,r_\mathrm{ISCO}$. Since $\dot M_*^\mathrm{in} \propto r^b$ where $b \approx 0.5$--0.7 (see below), we may expect that $\Delta M_*/M_*$ is also proportional to $r_\mathrm{ISCO}^b$ under the condition that the mass injection at large radii is approximately identical (irrespective of the value of $\chi$). In Fig.~\ref{fig4} we plot $A\,\hat r_\mathrm{ISCO}^{0.7}$ where $A=0.0325$ and 0.0162 for ${\hat r}_{\rm out}=10^3$ and $10^4$, respectively. It is found that these curves work very well as a fitting formula. Therefore, for a given density profile, we can say that $\Delta M_*/M_*$ is approximately proportional to $\ell^{-1}\,r_\mathrm{ISCO}^{0.7}$ (unless $\hat\ell$ is very small). Note that the power of $0.7$ is slightly larger than $b$. The reason for this is that (i) the value of $b$ is slightly larger for higher values of $\chi$ and (ii) the capture radius is not exactly $2\,r_\mathrm{ISCO}$ but it is slightly smaller and larger for higher and lower values of $\chi$, respectively; e.g., for $\chi=0.95$, $r_\mathrm{cap} \alt 2\,r_\mathrm{ISCO}$ but for $\chi=0$, $r_\mathrm{cap} \agt 2\,r_\mathrm{ISCO}$ (cf. also Figs.~\ref{fig7} and \ref{fig8}). 

Figures~\ref{fig3} and \ref{fig4} also indicate that $\Delta M_*/M_*$ depends on $\hat r_\mathrm{out}$ for a given value of $\hat\ell$. For the change from $\hat r_\mathrm{out}=10^3$ to $10^4$, it becomes about half. Thus $\Delta M_*/M_*$ may be written approximately as $\hat A \hat \ell^{-1} \,\hat r_\mathrm{ISCO}$ with $\hat A$ being a factor of order 0.1 that depends on the initial density distribution of the torus (for the tori studied in this paper $\hat A \approx 0.33$ and $0.16$ for $\hat r_\mathrm{out}=10^3$ and $10^4$).

Next, we discuss the dependence of $\Delta M_*/M_*$ on the viscous parameter $\alpha_\mathrm{vis}$. The upper panels of Fig.~\ref{fig5} plot $\Delta M_*/M_*$ as a function of time for $\hat \ell=10$ and $\hat r_\mathrm{out}=10^4$ with $\alpha_\mathrm{vis}=0.01$, 0.02, 0.05, 0.10, and $\chi=0.8$ (left, solid), with $\alpha_\mathrm{vis}=0.01$, 0.02, 0.05, and $\chi=0$ (left, dashed) and for $\hat\ell=6$ and $\chi=0.8$ with $\alpha_\mathrm{vis}=0.01$, 0.02, 0.05, and $\hat r_\mathrm{out}=10^4$ (right, solid) and $10^3$ (right, dashed). For $\alpha_\mathrm{vis}=0.1$, the value is typically about twice as large as that for $\alpha_\mathrm{vis}\leq 0.05$. A high-mass infall rate for the high value of $\alpha_\mathrm{vis}$ is found in the early stage of the torus evolution, i.e., by an initial impact of the viscous effect, and thus, it may be an artifact due to the initial setting. By contrast, for $\alpha_\mathrm{vis} \leq 0.05$, the total mass swallowed into the black hole depends only weakly on the viscous parameter, irrespective of $\hat\ell$ and $\hat r_\mathrm{out}$, although the infalling timescale depends on it. Thus, for $\alpha_\mathrm{vis} \leq 0.05$, the results do not appear to be affected by the initial artificial setting, but they are determined by the initial configuration of the torus. 

For smaller values of $\alpha_\mathrm{vis}$, the matter falls into the black hole spending a longer timescale. This is simply because the viscous timescale is longer for smaller values of $\alpha_\mathrm{vis}$. 
The lower panels of Fig.~\ref{fig5} plot the ejecta mass, $M_\mathrm{eje}$, as a function of time for the same models of the upper panels. These plots clearly show that for smaller values of $\alpha_\mathrm{vis}$, the mass ejection timescale (i.e., viscous timescale) is longer, although the final ejecta mass depends only weakly on $\alpha_\mathrm{vis}$. 

Figure~\ref{fig7} shows the same as the bottom panels of Fig.~\ref{fig25} but for $\hat \ell=10$, $\chi=0.8$, $\hat r_\mathrm{out}=10^4$, and $\alpha_\mathrm{vis}=0.01$ (top), 0.02 (middle), and 0.05 (bottom). The dashed slope is $\propto r^{0.7}$ for the top panel and $r^{1/2}$ for the middle and bottom panels. Again, we find that $\dot M_*(r)$ and $\dot M_*^\mathrm{in}(r)$ agree with each other (i.e., no outflow) inside the capture radius of $r \alt r_\mathrm{cap} \approx 2\,r_\mathrm{ISCO} \approx 5.8\,r_\mathrm{g}$, and for $r \agt r_\mathrm{cap}$, $\dot M_*^\mathrm{in}(r)$ increases with $r$. We can clearly identify that for $r_\mathrm{cap} \alt r \alt 100\,r_\mathrm{g}$ with  $\alpha_\mathrm{vis}=0.01$ or 0.02, $\dot M_*^\mathrm{in}$ is approximately proportional to $r^b$ where $b \approx 0.7$ and $1/2$ for $\alpha_\mathrm{vis}=0.01$ and 0.02, respectively. This behavior is consistent with the previous finding~\cite{2022ApJ...934..132H,2024MNRAS.532.4826T}. For $\alpha_\mathrm{vis}=0.05$, the dependence of $\dot M_*^\mathrm{in}$ is not well described by a power law, although the average increase rate is roughly proportional to $r^{1/2}$.

Figure~\ref{fig8} plots the same figure as the top panel of Fig.~\ref{fig7} (i.e., $\alpha_\mathrm{vis}=0.01$, $\hat\ell=10$, and $\hat r_\mathrm{out}=10^4$) but with $\chi=-0.8$ (top), $0$ (middle), and $0.95$ (bottom). Again the slope of $\dot M_*^\mathrm{in}$ for $r_\mathrm{cap} \alt r \alt 100\,r_\mathrm{g}$ is approximately written as $r^b$ with $b=0.6$--0.7 for $t \geq 10^5GM_\mathrm{BH}$. Because $r_\mathrm{ISCO}$ is smaller, the mass accretion rate on the horizon is lower for the larger values of $\chi$ (note also that the mass injection rate at $r \sim 200\,r_\mathrm{g}$ depends only weakly on $\chi$). As we already discussed, this fact primarily determines the strong dependence of $\Delta M_*/M_* (\propto r_\mathrm{ISCO}^{0.7})$ on $\chi$. 

\subsection{Ejecta velocity} \label{sec4C}
%%%%% VELOCITY

\begin{figure}
\includegraphics[width=0.49\textwidth]{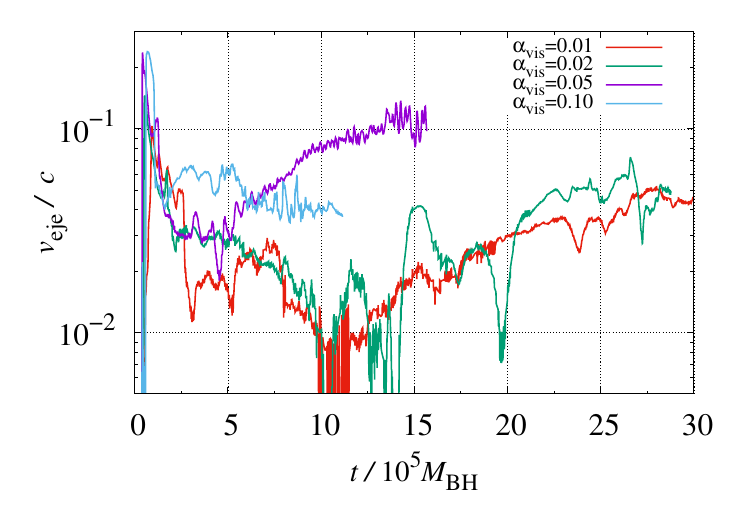}\\
\vspace{-5mm}
\includegraphics[width=0.49\textwidth]{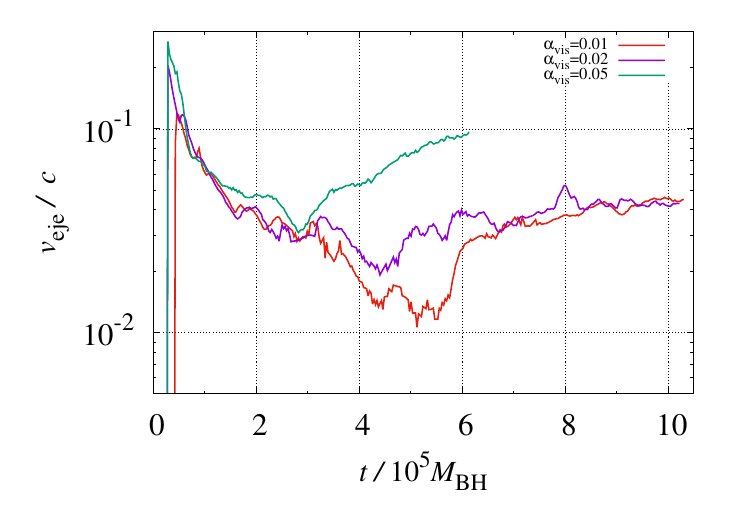}\\
\vspace{-5mm}
\includegraphics[width=0.49\textwidth]{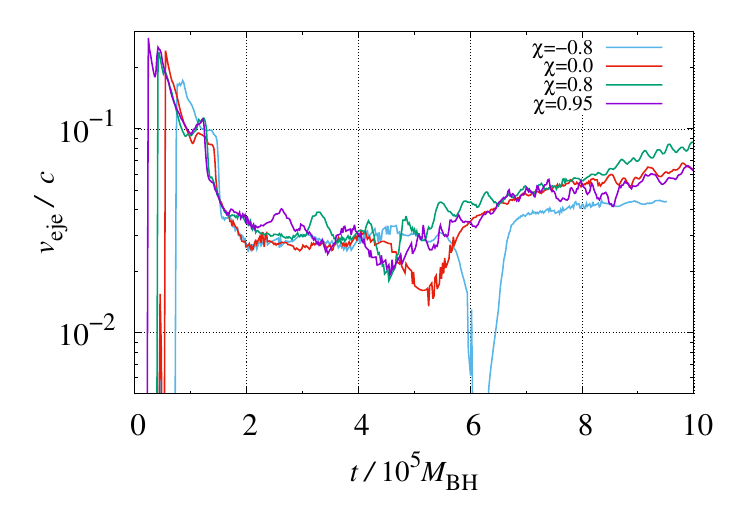}
\vspace{-7mm}
\caption{$v_\mathrm{eje}$ as a function of time for $\hat \ell=10$ and $\hat r_\mathrm{out}=10^4$ with $\alpha_\mathrm{vis}=0.01$, 0.02, 0.05, and 0.10 (top panel) and for $\hat\ell=6$ and $\hat r_\mathrm{out}=10^3$ with $\alpha_\mathrm{vis}=0.01$, 0.02, and 0.05 (middle panel). For both models $\chi=0.8$. The bottom panel shows the results for $\chi=-0.8$, 0, 0.8, and 0.95 with $\alpha_\mathrm{vis}=0.05$ and $\hat r_\mathrm{out}=10^4$.
}
\label{fig6}
\end{figure}

Figure~\ref{fig6} shows the average ejecta velocity, $v_\mathrm{eje}$, as a function of time for $\hat\ell=10$ and $\hat r_\mathrm{out}=10^4$ with $\alpha_\mathrm{vis}=0.01, 0.02, 0.05$, and $0.10$ (top), for $\hat \ell=6$ and $\hat r_\mathrm{out}=10^3$ with $\alpha_\mathrm{vis}=0.01$, 0.02, and 0.05 (middle), and for $\chi=0$, $0.8$, and $0.95$ with $\alpha_\mathrm{vis}=0.05$, $\hat\ell=10$, and $\hat r_\mathrm{out}=10^4$ (bottom). For the top and middle panels, $\chi=0.8$. 
The ejecta velocity, $v_\mathrm{eje}$, is initially high $\sim 0.1$--$0.2c$ and then relaxes approximately to 0.01--$0.05c$. The early high velocity is associated with the initial infall of a substantial fraction of matter and a small amount of surrounding matter around the torus. A part of the high-velocity matter is ejected toward the polar region and the magnitude of $\sim 0.2c$ is in good agreement with the X-ray observational result~\cite{Parker:2017wnh}. The relaxed velocity depends on the value of $\alpha_\mathrm{vis}$, and broadly speaking it is higher for larger values of $\alpha_\mathrm{vis}$. The resultant typical kinetic energy of the ejecta is $\sim 0.05$--0.1\% of the rest mass energy of the initial mass, $M_*c^2$, unless $\hat \ell$ is very small; for $M_*=M_\odot$, it is $\sim 10^{51}$\,erg. Here, for larger values of $\chi$ and for the larger values of $\alpha_\mathrm{vis}$, the kinetic energy of the ejecta is slightly larger. 

As discussed in Sec.~\ref{sec4.1} the mass ejection is driven primarily for the region of $r \agt r_\mathrm{c}$, although the energy injection resulting from the viscous heating is most efficient for $r \alt r_\mathrm{cap} \approx 2r_\mathrm{ISCO}$. This provides a schematic picture that the total viscous heating energy of $\sim GM_\mathrm{BH}\Delta M_*/(2r_\mathrm{cap})$ is converted to the kinetic energy of the ejecta,  $M_\mathrm{eje}v_\mathrm{eje}^2/2$. This leads to 
\begin{eqnarray}
v_\mathrm{eje} \sim (GM_\mathrm{BH}/r_\mathrm{cap})^{1/2}(\Delta M_*/M_\mathrm{eje})^{1/2}.  \label{eq26}
\end{eqnarray}
Here, $(GM_\mathrm{BH}/r_\mathrm{cap})^{1/2}=O(0.1c)$ and $(\Delta M_*/M_\mathrm{eje})^{1/2}=O(0.1)$ for $\hat \ell \agt 6$, and hence, the expected average ejecta velocity is of order $0.01c$. We note that for higher values of $\chi$, $(GM_\mathrm{BH}/r_\mathrm{cap})^{1/2}$ can be larger while $(\Delta M_*/M_\mathrm{eje})^{1/2}$ is smaller as we found in this paper. This estimate for the average velocity agrees broadly with the results shown in Fig.~\ref{fig6}. In reality, the viscous heating energy is also used to overcome the gravitational potential energy of the matter to be ejected, and thus, the velocity should be smaller than that obtained by the simple formula employed here. 

For $\alpha_\mathrm{vis}=0.01$ the ejecta velocity becomes lower than $0.01c$ or cannot be defined for some time span because the matter is bound in such a stage. This reflects the fact that a fraction of the matter, which initially moves outward, falls toward the black hole. However, after the continuous viscous heating near the black hole, such matter is eventually ejected as Fig.~\ref{fig5} indicates. 

In the late stage, the ejecta velocity tends to increase. The reason for this is that (i) the viscous heating, which is always more efficient near the black hole, is induced primarily by a blob of the infalling matter, (ii) the matter density around the black hole becomes smaller in the late stage, and hence, the impact of the viscous heating is strong enough to accelerate the matter to a high velocity (cf.~Eq.~\eqref{eq26}). Since the mass of the late-stage ejecta component is small, this does not contribute a lot to the total kinetic energy of the ejecta. However, such a high-velocity component may generate a shock at a collision with the pre-ejected matter, modifying the overall profile of the ejecta. 

The bottom panel of Fig.~\ref{fig6} compares the average ejecta velocity for $\chi=-0.8$, 0, 0.8, and 0.95 with $\alpha_\mathrm{vis}=0.05$ and $\hat r_\mathrm{out}=10^4$. We note that for $\chi=-0.8$ the ejecta component is absent around $t\approx 6 \times 10^5 GM_\mathrm{BH}$, and thus, $v_\mathrm{eje}$ is not defined. This figure shows that the velocity is slightly higher for the higher values of $\chi$ reflecting the fact that the mass ejection is dominantly induced around $r=2\,r_\mathrm{cap}$, which is smaller for larger values of $\chi$. However, the effect by the black hole spin is not as appreciable as that of the viscous coefficient, $\alpha_\mathrm{vis}$.

\subsection{Spin up of black holes}

%%%%%%%% SPIN UP BH

\begin{figure}
\hspace{-6mm}
\includegraphics[width=0.51\textwidth]{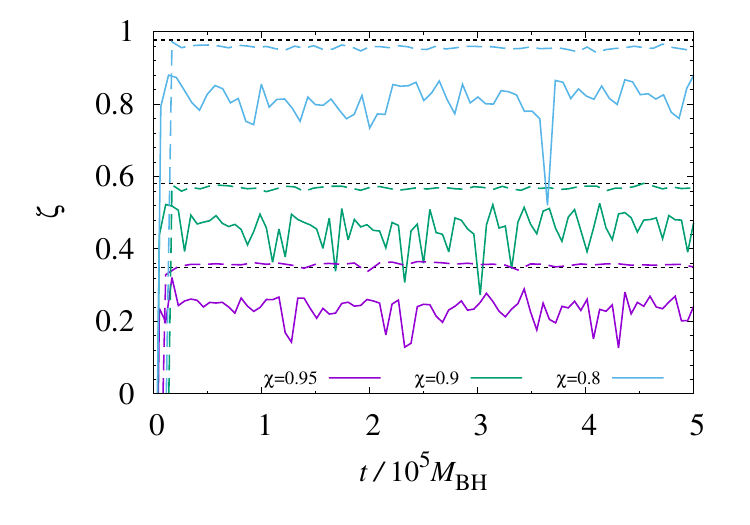}
\vspace{-10mm}
\caption{The evolution of the $\zeta$ parameter for $\chi=0.8$, 0.9, and 0.95 with $\hat \ell=10$, $\hat r_\mathrm{out}=10^4$, and $\alpha_\mathrm{vis}=0.02$ (dashed curves) and 0.05 (solid curves). We note that smoothing is operated for plotting the $\zeta$ parameter because the raw curves are very oscillatory. The dotted horizontal lines denote $\zeta=\zeta_\mathrm{ISCO}$ for each spin. 
}
\label{fig9}
\end{figure}

Figure~\ref{fig9} shows the $\zeta$ parameter as a function of time for $\chi=0.8$, 0.9, and 0.95 with $\hat \ell=10$ and $\hat r_\mathrm{out}=10^4$. The dashed horizontal lines show the values expected for the case that the matter adiabatically falls into the black holes from the innermost stable circular orbit, i.e., $\zeta=\zeta_\mathrm{ISCO}$. We find that for $\alpha_\mathrm{vis}=0.02$, the curves are quite close to the dashed lines, indicating that the matter falls into the black hole from the vicinity of the innermost stable circular orbits, approximately preserving the specific energy and angular momentum there. For the higher value of $\alpha_\mathrm{vis}$ as $0.05$, the $\zeta$ values are smaller than those for $\alpha_\mathrm{vis}=0.02$, indicating that the angular momentum transport process is more efficient near the innermost stable circular orbits. However, even with $\alpha_\mathrm{vis}=0.05$ and $\chi=0.95$, the $\zeta$ value is positive; i.e., the black hole spins up by the matter accretion. This result is in contrast to those in magnetohydrodynamics simulations~\cite{2004ApJ...602..312G,2022MNRAS.511.3795N}, in which not the spin-up but the spin-down is concluded for highly spinning black holes. The reason for this is that in magnetohydrodynamics simulations, not only the matter infall by the (effectively) viscous effect but also by the additional magnetohydrodynamics processes such as the Blandford-Znajek one \cite{Blandford1977}, which plays a role in reducing the black hole spin, determine the evolution of the black hole spin. 

\subsection{Dependence of $\Delta M_*$ on $\Gamma$} \label{sec4E}

\begin{figure}
\hspace{-6mm}
\includegraphics[width=0.51\textwidth]{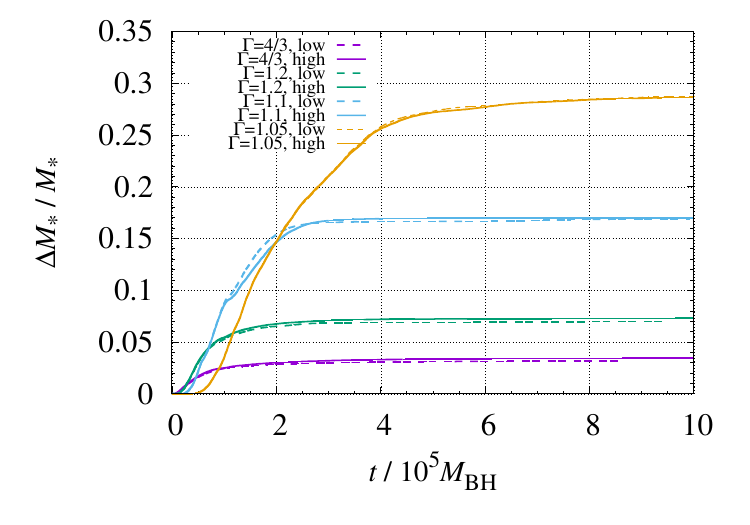}
\vspace{-10mm}
\caption{Evolution of $\Delta M_*/M_*$ for $\hat\ell=10$, $\hat r_\mathrm{out}=10^4$, $\chi=0.8$, and $\Gamma=4/3$, 1.2, 1.1, and 1.05. The results with two grid resolutions (solid curves are better resolution results) are shown. 
}
\label{fig10}
\end{figure}

Figure~\ref{fig10} shows the evolution of $\Delta M_*/M_*$ for $\chi=0.8$, $\hat \ell=10$, and $\hat r_\mathrm{out}=10^4$  with $\Gamma=4/3$, 1.2, 1.1, and 1.05. The solid and dashed curves denote the results with high and standard grid resolutions, respectively. It is found that for the smaller values of $\Gamma$, the fraction of the matter swallowed into the black hole increases because the viscous and subsequent shock heating efficiency are lower for the lower values of $\Gamma$, and as a result, mass ejection is suppressed. This tendency is, in particular, remarkable for $\Gamma \rightarrow 1$. This result clearly reflects that the mass ejection is driven by the viscous and subsequent shock heating effects. We performed the simulations also for $\chi=0$ and found essentially the same tendency ($\Delta M_*/M_*$ slightly shifts upward systematically).

We also analyse $\dot M_*(r)$ and $\dot M_*^\mathrm{in}(r)$ for $\Gamma < 4/3$ and find that irrespective of $\Gamma$ the capture radius is located at $\approx 2\,r_\mathrm{isco}$. This indicates that the capture radius is determined purely by the spacetime structure but not the heating (and cooling) efficiency. 
By contrast the slope of $M_*^\mathrm{in}$ for $r_\mathrm{cap} \alt r \alt r_\mathrm{c}$ becomes gentler for $\Gamma < 4/3$. This is as expected, because $\Delta M_*/M_*$ is larger for smaller values of $\Gamma$ while the value of $r_\mathrm{cap}$ is fixed. 
The comparison between the results with the two grid resolutions (shown in Fig.~\ref{fig10}) illustrates a good convergence of the numerical results for the employed grid resolutions. 

\section{Discussion}\label{sec5}

\subsection{Increase of supermassive black hole mass}

Here, we consider a very simple toy model for the growth of an SMBH by the accretion of matter, which comes from a tidal disruption of ordinary stars. For simplicity, we employ the mass and radius of the solar-type star as an example. The tidal disruption can happen if the periastron is smaller than the tidal radius $r_\mathrm{t}$, defined in Eq.~\eqref{eq2}. 
Since $r_\mathrm{t}$ has to be larger than the horizon radius $r_\mathrm{H}$, we here suppose that the black hole mass is smaller than $\sim 10^8M_\odot$. 

Assuming that the tidal disruption typically happens at $r=r_\mathrm{t}$ and the resulting tidal debris eventually forms an accretion disk of a nearly circular orbit, the specific angular momentum of the disk may be approximately written as $\ell \approx \sqrt{GM_\mathrm{BH}r_\mathrm{t}}$. In this paper, we found that for such a disk/torus, the fraction $\xi=\Delta M_*/M*=\hat A \hat\ell^{-1} \,\hat r_\mathrm{ISCO}$ can fall into the black hole.  
For $\chi=0$, we find $\xi=\xi_0 \hat \ell^{-1}$ with $\xi_0 \sim 0.6$--1.1, and in the following, we employ the value for $\chi=0$ as $\xi_0 \sim 1$. 

Let $N$ be the tidal disruption rate. Then the evolution equation of the black hole mass may be written as
\beqn
\dot M_\mathrm{BH}=N \xi M_\star. \label{eq4.1}
\eeqn
Here $N$ should depend on the mass of the SMBHs~\cite{1999MNRAS.306...35S, 1999MNRAS.309..447M}; for lower-mass black holes, $N$ could be higher in the present-day galaxies. For simplicity, we bravely analyze Eq.~(\ref{eq4.1}) assuming the form of $N=N_0(M_\mathrm{BH}/M_\mathrm{BH,0})^{-\alpha}$ where $N_0$ and $\alpha$ are constants, and $M_\mathrm{BH,0}$ denotes the initial value of $M_\mathrm{BH}$. 

Using Eq.~(\ref{eq2}), Eq.~(\ref{eq4.1}) is written as
\beqn
\dot m_\mathrm{BH}=\xi_0 N_0 m_\mathrm{BH,0} ^\alpha m_\mathrm{BH}^{1/3-\alpha} 
\left({GM_\star \over R_\star}\right)^{1/2},
\eeqn
where $m_\mathrm{BH}=M_\mathrm{BH}/M_\star$, i.e., a dimensionless quantity, with $m_\mathrm{BH,0}$ its initial value, and then, we obtain
\beqn
m_\mathrm{BH}=m_\mathrm{BH,0}\left(1+ {t \over t_N}\right)^{3/(2+3\alpha)},
\label{eq7}
\eeqn
where
\beqn
t_N={3 \over (2+3\alpha)\xi_0 N_0} m_\mathrm{BH,0}^{2/3}\left({R_\star \over GM_\star}\right)^{1/2}. 
\eeqn
For $\alpha=0$, $M_\star=M_\odot$, and $R_\star=R_\odot$, $t_N$ is written as
\beqn
t_N &\approx& 2.2\times 10^8 \,\mathrm{yr} \, \xi_0^{-1}
\left({N_0 \over 10^{-2}\,\mathrm{yr}^{-1}}\right)^{-1}
\left({m_\mathrm{BH,0} \over 10^5}\right)^{2/3} \label{eq9},~~~
\eeqn
and with larger values of $\alpha > 0$, the timescale becomes shorter. Equation~\eqref{eq9} suggests that for a relatively less massive SMBH, the growth timescale can be shorter than $5 \times 10^8$\,yr, which is approximately equal to the age of the universe at the cosmological redshift $z=10$ if the infall rate of stars that can be tidally disrupted is high enough $\agt 10^{-2}$/yr (i.e., the mass infall rate is $10^{-2}M_\star$/yr). The required rate is much higher than the tidal disruption rate in the present-day galaxies with an SMBH of mass $\sim 10^6M_\odot$~\cite{1999MNRAS.306...35S, 1999MNRAS.309..447M} but such a mass accretion rate is often assumed for the formation of supermassive stars leading to a massive seed of SMBHs in the early universe~\cite{2013ApJ...778..178H}. 

If the tidal disruption rate depends only weakly on the black hole mass and $\alpha$ is less than $1/3$, the growth of the black hole mass can be accelerated for $t > t_N$ as Eq.~\eqref{eq7} indicates. By contrast, if the infall rate of the stars decreases with the increase of the black hole mass, the mass increase by the tidal disruption events would be saturated. If so, other mechanisms, such as rapid gas accretion, are necessary to increase the mass of the SMBH beyond $\sim 10^6M_\odot$. 
Equation~\eqref{eq9} also indicates that for SMBHs with mass $\agt 10^7M_\odot$, the tidal disruption and subsequent mass accretion might not be an efficient mechanism for the rapid mass growth in the early universe because $t_N$ is likely to be longer than the age of the universe at high redshifts (unless $N_0$ is extra-ordinary high). 

For the present-day universe, the tidal disruption rate is typically smaller than $10^{-4}$/yr~\cite{1999MNRAS.306...35S, 1999MNRAS.309..447M}. Equation~\eqref{eq9} then suggests that the growth timescale of the SMBH associated with the tidal disruption is longer than the Hubble time $\approx 1.38 \times 10^{10}$\,yr for $M_\mathrm{BH} \leq 2 \times 10^5M_\mathrm{BH}$. Thus, the tidal disruption events might not contribute to the recent growth of low-mass SMBHs. 

\subsection{Energy injection from ejecta}

In this subsection, we do not always focus on torus formation by tidal disruption but simply assume that an SMBH grows via mass accretion from tori. 
In this paper, we find that most of the torus matter is ejected from the system with an average ejecta velocity of a few percent of the speed of light. The typical ejecta mass is about one order of magnitude larger than the mass of the matter swallowed into the black hole. This implies that if the black hole of mass  $M_\mathrm{BH}$ has grown from a seed of mass $M_\mathrm{BH,0} \ll M_\mathrm{BH}$, the total ejecta mass would be $\sim 10M_\mathrm{BH}$ with the total kinetic energy of $10^{-2}M_\mathrm{BH}c^2$; e.g., for $M_\mathrm{BH}=10^7M_\odot$, the kinetic energy is of order $10^{59}$\,erg. The ejecta would subsequently interact with the interstellar gas in the corresponding galaxies and dissipate the kinetic energy, heating up the interstellar matter. Thus, in the early stage of the galaxy formation in which an SMBH grows rapidly, a significant kinetic energy injection to the surrounding environment should accompany it. 

If the temperature of the interstellar gas exceeds $\sim 10^4$\,K, cooling by the bremsstrahlung will proceed (e.g., Ref.~\cite{1986RvMP...58....1S}). Assuming that the hydrogen gas number density is higher than $1\,\mathrm{cm^{-3}}$, the cooling timescale is shorter than the age of the universe at the redshift 10 ($\sim 5\times 10^8$\,yr). This implies that the cooling is efficient. Then, assuming that matter accretion continues for the entire phase of the SMBH growth, the average luminosity by this cooling may be estimated by $L\sim 10^{-2}M_\mathrm{BH}/\tau$ where $\tau$ denotes the age of the universe at the time that the SMBH mass is $M_\mathrm{BH}$, and thus,
\beqn
L \sim 10^{43}\,\mathrm{erg/s}\left({M_\mathrm{BH} \over 10^7M_\odot}\right)
\left({\tau \over 5 \times 10^8\,\mathrm{yr}}\right)^{-1}.  \label{eq33}
\eeqn
This is the average value, and in the enhanced growth epoch of an SMBH for which $\tau$ is short (i.e., the mass accretion rates are intermittently high), the luminosity may be much higher.
Equation~\eqref{eq33} shows that the average luminosity by this process is lower than the Eddington luminosity, $L_\mathrm{Edd} \approx 1.4 \times 10^{45}\,\mathrm{erg/s} (M_\mathrm{BH}/10^7M_\odot)$, with which the super-Eddington accretion disk is likely to radiate in the accretion phase, but $L$ is still comparable to the luminosity of the present-day galaxies.

We note that our present results on the ejecta mass are based strongly on the assumption of the absence of cooling effects via photon emission during the accretion and mass ejection processes. In the presence of radiative cooling, the mass ejection is suppressed because the thermal energy generated by the viscous effects should be consumed by the photon emission. However, the result of this paper indicates that before such a stage occurs, a majority of the matter of the torus initially present is likely to be ejected if the initial state of the torus is dense. Thus, the estimate of $L$ here would be a reasonable order of the magnitude of the luminosity resulting from the kinetic energy.

\section{Summary}\label{sec6}

We performed viscous hydrodynamics simulations for tori orbiting spinning black holes with large typical radii in general relativity.  
The simulations were performed for a long timescale of order $10^6GM_\mathrm{BH}$, which is by one order of magnitude longer than the typical timescales for cutting-edge radiation magnetohydrodynamics simulations for accretion flows~(e.g., Refs.~\cite{2014MNRAS.439..503S, 2014MNRAS.441.3177M, 2016MNRAS.456.3929S, 2022MNRAS.511.3795N}) and for a variety of the specific angular momentum of the tori and dimensionless spin of the black hole. 
In this work, we did not take into account the cooling effects via photon emission because we paid attention to a photon-trapped dense torus, which is likely to be formed after tidal disruption of ordinary stars by SMBHs or a rapid mass inflow. 

We made the following findings: 
\begin{itemize}
\item The fraction of the rest mass of a torus that falls into the black hole, $\Delta M_*/M_*$, is approximately proportional to $\ell^{-1}$ irrespective of the black hole spin for $\hat\ell \agt 6$. 
\item Irrespective of the initial setting, the majority of the torus matter is ejected from the system, unless the specific angular momentum of the torus is close to $\ell_\mathrm{ISCO}$. 
\item For high dimensionless spins with $\chi \geq 0.8$, the fraction of the mass that falls into the black hole is often less than 10\%, while more than 90\% of the mass is ejected from the system. Depending on the dimensionless spin, the fraction of the mass infall into the black hole changes by a factor of several. 
\item The mass outflow is driven from a region of $r \agt r_\mathrm{cap}$ where the capture radius $r_\mathrm{cap}$ is $\sim 2\,r_\mathrm{ISCO}$. $r_\mathrm{cap}$ is determined purely by the general relativistic gravity. For $r \alt r_\mathrm{cap}$, the mass outflow is absent, i.e., the matter inside the capture radius $r_\mathrm{cap}$ falls into the black hole. This implies that the general relativistic effect is essential in this problem. 
\item The gross mass infall rate $\dot M_*^\mathrm{in}$ is approximately proportional to $r^b$ with $b=0.5$--0.7 for $r_\mathrm{cap} \leq r \alt r_\mathrm{c}$. The value of $b$ can depend on the equation of state and cooling efficiency (e.g., Ref.~\cite{2002ApJ...565.1101A}). 
\item Associated with the presence of the capture radius and the relation of $\dot M_*^\mathrm{in} \propto r^b$, the spin-dependence of the fraction of the mass infall, $\Delta M_*/M_*$, is approximately proportional to $r_\mathrm{ISCO}^{0.7}$. Therefore, $\Delta M_*/M_*$ is approximately written as $\hat A \hat \ell^{-1} \hat r_\mathrm{ISCO}^{0.7}$ for a given density profile of the torus with $\hat \ell \agt 6$. Here $\hat A$ is a constant of order 0.1, which depends on the initial profile of the torus.
\item The velocity of the ejecta is typically a few percent of the speed of light in the present setting which gives $(\Delta M_*/M_\mathrm{eje})^{1/2}\sim 0.1$--0.3. For the larger values of $\alpha_\mathrm{vis}$, the velocity is slightly higher. The resultant typical kinetic energy of the ejecta is $\sim 0.1$\% of the rest mass energy of the initial mass.
\item Black holes do not spin down in viscous hydrodynamics at least for $\chi \leq 0.95$. This result is in clear contrast to those in magnetohydrodynamics \cite{2004ApJ...602..312G, 2022MNRAS.511.3795N}, in which the Blandford-Znajek mechanism~\cite{Blandford1977} can significantly contribute to the spin down at $\chi \sim 0.95$. 
\end{itemize}
Among these findings, the approximate fitting formula, $\Delta M_*/M_* \propto \hat \ell^{-1} \hat r_\mathrm{ISCO}^{0.7}$, will be useful for modeling the growth of the SMBH mass. 

In this paper, we did not take into account the radiation transfer effects. In the late phase of the accretion, the density of the torus becomes low enough to shorten the diffusion timescale of photons, which will be shorter than the viscous timescale. In such a phase, the cooling by the photon emission plays an important role. In the presence of efficient cooling, the thermal energy generated by the viscous heating is consumed by the cooling, and as a result, the mass ejection would be suppressed, and the matter accretion onto the black hole would be enhanced. However, the result of this paper indicates that before such a stage comes, a majority of the initial torus matter in which photons are trapped is likely to be ejected with the ejection velocity of a few percent of the speed of light. 

\acknowledgements

We thank Kenta Hotokezaka, Kunihito Ioka, Kohta Murase, and Alexis Reboul-Salze for useful discussions and valuable comments. This work was in part supported by Grant-in-Aid for Scientific Research (grant No.~23H04900) of Japanese MEXT/JSPS.

\bibliographystyle{apsrev4-2}
\bibliography{reference}

\begin{thebibliography}{49}%
\makeatletter
\providecommand \@ifxundefined [1]{%
 \@ifx{#1\undefined}
}%
\providecommand \@ifnum [1]{%
 \ifnum #1\expandafter \@firstoftwo
 \else \expandafter \@secondoftwo
 \fi
}%
\providecommand \@ifx [1]{%
 \ifx #1\expandafter \@firstoftwo
 \else \expandafter \@secondoftwo
 \fi
}%
\providecommand \natexlab [1]{#1}%
\providecommand \enquote  [1]{``#1''}%
\providecommand \bibnamefont  [1]{#1}%
\providecommand \bibfnamefont [1]{#1}%
\providecommand \citenamefont [1]{#1}%
\providecommand \href@noop [0]{\@secondoftwo}%
\providecommand \href [0]{\begingroup \@sanitize@url \@href}%
\providecommand \@href[1]{\@@startlink{#1}\@@href}%
\providecommand \@@href[1]{\endgroup#1\@@endlink}%
\providecommand \@sanitize@url [0]{\catcode `\\12\catcode `\$12\catcode
  `\&12\catcode `\#12\catcode `\^12\catcode `\_12\catcode `\%12\relax}%
\providecommand \@@startlink[1]{}%
\providecommand \@@endlink[0]{}%
\providecommand \url  [0]{\begingroup\@sanitize@url \@url }%
\providecommand \@url [1]{\endgroup\@href {#1}{\urlprefix }}%
\providecommand \urlprefix  [0]{URL }%
\providecommand \Eprint [0]{\href }%
\providecommand \doibase [0]{https://doi.org/}%
\providecommand \selectlanguage [0]{\@gobble}%
\providecommand \bibinfo  [0]{\@secondoftwo}%
\providecommand \bibfield  [0]{\@secondoftwo}%
\providecommand \translation [1]{[#1]}%
\providecommand \BibitemOpen [0]{}%
\providecommand \bibitemStop [0]{}%
\providecommand \bibitemNoStop [0]{.\EOS\space}%
\providecommand \EOS [0]{\spacefactor3000\relax}%
\providecommand \BibitemShut  [1]{\csname bibitem#1\endcsname}%
\let\auto@bib@innerbib\@empty
%</preamble>
\bibitem [{\citenamefont {{Fan}}\ \emph {et~al.}(2023)\citenamefont {{Fan}},
  \citenamefont {{Ba{\~n}ados}},\ and\ \citenamefont {{Simcoe}}}]{Fan2023aug}%
  \BibitemOpen
  \bibfield  {author} {\bibinfo {author} {\bibfnamefont {X.}~\bibnamefont
  {{Fan}}}, \bibinfo {author} {\bibfnamefont {E.}~\bibnamefont
  {{Ba{\~n}ados}}},\ and\ \bibinfo {author} {\bibfnamefont {R.~A.}\
  \bibnamefont {{Simcoe}}},\ }\href
  {https://doi.org/10.1146/annurev-astro-052920-102455} {\bibfield  {journal}
  {\bibinfo  {journal} {\araa}\ }\textbf {\bibinfo {volume} {61}},\ \bibinfo
  {pages} {373} (\bibinfo {year} {2023})},\ \Eprint
  {https://arxiv.org/abs/2212.06907} {arXiv:2212.06907 [astro-ph.GA]}
  \BibitemShut {NoStop}%
\bibitem [{\citenamefont {{Harikane}}\ \emph {et~al.}(2023)\citenamefont
  {{Harikane}}, \citenamefont {{Zhang}}, \citenamefont {{Nakajima}},
  \citenamefont {{Ouchi}}, \citenamefont {{Isobe}}, \citenamefont {{Ono}},
  \citenamefont {{Hatano}}, \citenamefont {{Xu}},\ and\ \citenamefont
  {{Umeda}}}]{2023ApJ...959...39H}%
  \BibitemOpen
  \bibfield  {author} {\bibinfo {author} {\bibfnamefont {Y.}~\bibnamefont
  {{Harikane}}}, \bibinfo {author} {\bibfnamefont {Y.}~\bibnamefont {{Zhang}}},
  \bibinfo {author} {\bibfnamefont {K.}~\bibnamefont {{Nakajima}}}, \bibinfo
  {author} {\bibfnamefont {M.}~\bibnamefont {{Ouchi}}}, \bibinfo {author}
  {\bibfnamefont {Y.}~\bibnamefont {{Isobe}}}, \bibinfo {author} {\bibfnamefont
  {Y.}~\bibnamefont {{Ono}}}, \bibinfo {author} {\bibfnamefont
  {S.}~\bibnamefont {{Hatano}}}, \bibinfo {author} {\bibfnamefont
  {Y.}~\bibnamefont {{Xu}}},\ and\ \bibinfo {author} {\bibfnamefont
  {H.}~\bibnamefont {{Umeda}}},\ }\href
  {https://doi.org/10.3847/1538-4357/ad029e} {\bibfield  {journal} {\bibinfo
  {journal} {\apj}\ }\textbf {\bibinfo {volume} {959}},\ \bibinfo {eid} {39}
  (\bibinfo {year} {2023})},\ \Eprint {https://arxiv.org/abs/2303.11946}
  {arXiv:2303.11946 [astro-ph.GA]} \BibitemShut {NoStop}%
\bibitem [{\citenamefont {{Bogd{\'a}n}}\ \emph {et~al.}(2024)\citenamefont
  {{Bogd{\'a}n}}, \citenamefont {{Goulding}}, \citenamefont {{Natarajan}},
  \citenamefont {{Kov{\'a}cs}}, \citenamefont {{Tremblay}}, \citenamefont
  {{Chadayammuri}}, \citenamefont {{Volonteri}}, \citenamefont {{Kraft}},
  \citenamefont {{Forman}}, \citenamefont {{Jones}}, \citenamefont
  {{Churazov}},\ and\ \citenamefont {{Zhuravleva}}}]{Bogdan2024jan}%
  \BibitemOpen
  \bibfield  {author} {\bibinfo {author} {\bibfnamefont {{\'A}.}~\bibnamefont
  {{Bogd{\'a}n}}}, \bibinfo {author} {\bibfnamefont {A.~D.}\ \bibnamefont
  {{Goulding}}}, \bibinfo {author} {\bibfnamefont {P.}~\bibnamefont
  {{Natarajan}}}, \bibinfo {author} {\bibfnamefont {O.~E.}\ \bibnamefont
  {{Kov{\'a}cs}}}, \bibinfo {author} {\bibfnamefont {G.~R.}\ \bibnamefont
  {{Tremblay}}}, \bibinfo {author} {\bibfnamefont {U.}~\bibnamefont
  {{Chadayammuri}}}, \bibinfo {author} {\bibfnamefont {M.}~\bibnamefont
  {{Volonteri}}}, \bibinfo {author} {\bibfnamefont {R.~P.}\ \bibnamefont
  {{Kraft}}}, \bibinfo {author} {\bibfnamefont {W.~R.}\ \bibnamefont
  {{Forman}}}, \bibinfo {author} {\bibfnamefont {C.}~\bibnamefont {{Jones}}},
  \bibinfo {author} {\bibfnamefont {E.}~\bibnamefont {{Churazov}}},\ and\
  \bibinfo {author} {\bibfnamefont {I.}~\bibnamefont {{Zhuravleva}}},\ }\href
  {https://doi.org/10.1038/s41550-023-02111-9} {\bibfield  {journal} {\bibinfo
  {journal} {Nature Astronomy}\ }\textbf {\bibinfo {volume} {8}},\ \bibinfo
  {pages} {126} (\bibinfo {year} {2024})},\ \Eprint
  {https://arxiv.org/abs/2305.15458} {arXiv:2305.15458 [astro-ph.GA]}
  \BibitemShut {NoStop}%
\bibitem [{\citenamefont {{Goulding}}\ \emph {et~al.}(2023)\citenamefont
  {{Goulding}}, \citenamefont {{Greene}}, \citenamefont {{Setton}},
  \citenamefont {{Labbe}}, \citenamefont {{Bezanson}}, \citenamefont
  {{Miller}}, \citenamefont {{Atek}}, \citenamefont {{Bogd{\'a}n}},
  \citenamefont {{Brammer}}, \citenamefont {{Chemerynska}}, \citenamefont
  {{Cutler}}, \citenamefont {{Dayal}}, \citenamefont {{Fudamoto}},
  \citenamefont {{Fujimoto}}, \citenamefont {{Furtak}}, \citenamefont
  {{Kokorev}}, \citenamefont {{Khullar}}, \citenamefont {{Leja}}, \citenamefont
  {{Marchesini}}, \citenamefont {{Natarajan}}, \citenamefont {{Nelson}},
  \citenamefont {{Oesch}}, \citenamefont {{Pan}}, \citenamefont {{Papovich}},
  \citenamefont {{Price}}, \citenamefont {{van Dokkum}}, \citenamefont
  {{Wang}}, \citenamefont {{Weaver}}, \citenamefont {{Whitaker}},\ and\
  \citenamefont {{Zitrin}}}]{Goulding2023sep}%
  \BibitemOpen
  \bibfield  {author} {\bibinfo {author} {\bibfnamefont {A.~D.}\ \bibnamefont
  {{Goulding}}}, \bibinfo {author} {\bibfnamefont {J.~E.}\ \bibnamefont
  {{Greene}}}, \bibinfo {author} {\bibfnamefont {D.~J.}\ \bibnamefont
  {{Setton}}}, \bibinfo {author} {\bibfnamefont {I.}~\bibnamefont {{Labbe}}},
  \bibinfo {author} {\bibfnamefont {R.}~\bibnamefont {{Bezanson}}}, \bibinfo
  {author} {\bibfnamefont {T.~B.}\ \bibnamefont {{Miller}}}, \bibinfo {author}
  {\bibfnamefont {H.}~\bibnamefont {{Atek}}}, \bibinfo {author} {\bibfnamefont
  {{\'A}.}~\bibnamefont {{Bogd{\'a}n}}}, \bibinfo {author} {\bibfnamefont
  {G.}~\bibnamefont {{Brammer}}}, \bibinfo {author} {\bibfnamefont
  {I.}~\bibnamefont {{Chemerynska}}}, \bibinfo {author} {\bibfnamefont {S.~E.}\
  \bibnamefont {{Cutler}}}, \bibinfo {author} {\bibfnamefont {P.}~\bibnamefont
  {{Dayal}}}, \bibinfo {author} {\bibfnamefont {Y.}~\bibnamefont {{Fudamoto}}},
  \bibinfo {author} {\bibfnamefont {S.}~\bibnamefont {{Fujimoto}}}, \bibinfo
  {author} {\bibfnamefont {L.~J.}\ \bibnamefont {{Furtak}}}, \bibinfo {author}
  {\bibfnamefont {V.}~\bibnamefont {{Kokorev}}}, \bibinfo {author}
  {\bibfnamefont {G.}~\bibnamefont {{Khullar}}}, \bibinfo {author}
  {\bibfnamefont {J.}~\bibnamefont {{Leja}}}, \bibinfo {author} {\bibfnamefont
  {D.}~\bibnamefont {{Marchesini}}}, \bibinfo {author} {\bibfnamefont
  {P.}~\bibnamefont {{Natarajan}}}, \bibinfo {author} {\bibfnamefont
  {E.}~\bibnamefont {{Nelson}}}, \bibinfo {author} {\bibfnamefont {P.~A.}\
  \bibnamefont {{Oesch}}}, \bibinfo {author} {\bibfnamefont {R.}~\bibnamefont
  {{Pan}}}, \bibinfo {author} {\bibfnamefont {C.}~\bibnamefont {{Papovich}}},
  \bibinfo {author} {\bibfnamefont {S.~H.}\ \bibnamefont {{Price}}}, \bibinfo
  {author} {\bibfnamefont {P.}~\bibnamefont {{van Dokkum}}}, \bibinfo {author}
  {\bibfnamefont {B.}~\bibnamefont {{Wang}}}, \bibinfo {author} {\bibfnamefont
  {J.~R.}\ \bibnamefont {{Weaver}}}, \bibinfo {author} {\bibfnamefont {K.~E.}\
  \bibnamefont {{Whitaker}}},\ and\ \bibinfo {author} {\bibfnamefont
  {A.}~\bibnamefont {{Zitrin}}},\ }\href
  {https://doi.org/10.3847/2041-8213/acf7c5} {\bibfield  {journal} {\bibinfo
  {journal} {\apjl}\ }\textbf {\bibinfo {volume} {955}},\ \bibinfo {eid} {L24}
  (\bibinfo {year} {2023})},\ \Eprint {https://arxiv.org/abs/2308.02750}
  {arXiv:2308.02750 [astro-ph.GA]} \BibitemShut {NoStop}%
\bibitem [{\citenamefont {{Kov{\'a}cs}}\ \emph {et~al.}(2024)\citenamefont
  {{Kov{\'a}cs}}, \citenamefont {{Bogd{\'a}n}}, \citenamefont {{Natarajan}},
  \citenamefont {{Werner}}, \citenamefont {{Azadi}}, \citenamefont
  {{Volonteri}}, \citenamefont {{Tremblay}}, \citenamefont {{Chadayammuri}},
  \citenamefont {{Forman}}, \citenamefont {{Jones}},\ and\ \citenamefont
  {{Kraft}}}]{Kovacs2024apr}%
  \BibitemOpen
  \bibfield  {author} {\bibinfo {author} {\bibfnamefont {O.~E.}\ \bibnamefont
  {{Kov{\'a}cs}}}, \bibinfo {author} {\bibfnamefont {{\'A}.}~\bibnamefont
  {{Bogd{\'a}n}}}, \bibinfo {author} {\bibfnamefont {P.}~\bibnamefont
  {{Natarajan}}}, \bibinfo {author} {\bibfnamefont {N.}~\bibnamefont
  {{Werner}}}, \bibinfo {author} {\bibfnamefont {M.}~\bibnamefont {{Azadi}}},
  \bibinfo {author} {\bibfnamefont {M.}~\bibnamefont {{Volonteri}}}, \bibinfo
  {author} {\bibfnamefont {G.~R.}\ \bibnamefont {{Tremblay}}}, \bibinfo
  {author} {\bibfnamefont {U.}~\bibnamefont {{Chadayammuri}}}, \bibinfo
  {author} {\bibfnamefont {W.~R.}\ \bibnamefont {{Forman}}}, \bibinfo {author}
  {\bibfnamefont {C.}~\bibnamefont {{Jones}}},\ and\ \bibinfo {author}
  {\bibfnamefont {R.~P.}\ \bibnamefont {{Kraft}}},\ }\href
  {https://doi.org/10.3847/2041-8213/ad391f} {\bibfield  {journal} {\bibinfo
  {journal} {\apjl}\ }\textbf {\bibinfo {volume} {965}},\ \bibinfo {eid} {L21}
  (\bibinfo {year} {2024})},\ \Eprint {https://arxiv.org/abs/2403.14745}
  {arXiv:2403.14745 [astro-ph.GA]} \BibitemShut {NoStop}%
\bibitem [{\citenamefont {{Inayoshi}}\ \emph {et~al.}(2020)\citenamefont
  {{Inayoshi}}, \citenamefont {{Visbal}},\ and\ \citenamefont
  {{Haiman}}}]{Inayoshi2020aug}%
  \BibitemOpen
  \bibfield  {author} {\bibinfo {author} {\bibfnamefont {K.}~\bibnamefont
  {{Inayoshi}}}, \bibinfo {author} {\bibfnamefont {E.}~\bibnamefont
  {{Visbal}}},\ and\ \bibinfo {author} {\bibfnamefont {Z.}~\bibnamefont
  {{Haiman}}},\ }\href {https://doi.org/10.1146/annurev-astro-120419-014455}
  {\bibfield  {journal} {\bibinfo  {journal} {\araa}\ }\textbf {\bibinfo
  {volume} {58}},\ \bibinfo {pages} {27} (\bibinfo {year} {2020})},\ \Eprint
  {https://arxiv.org/abs/1911.05791} {arXiv:1911.05791 [astro-ph.GA]}
  \BibitemShut {NoStop}%
\bibitem [{\citenamefont {{Volonteri}}\ \emph {et~al.}(2021)\citenamefont
  {{Volonteri}}, \citenamefont {{Habouzit}},\ and\ \citenamefont
  {{Colpi}}}]{Volonteri2021sep}%
  \BibitemOpen
  \bibfield  {author} {\bibinfo {author} {\bibfnamefont {M.}~\bibnamefont
  {{Volonteri}}}, \bibinfo {author} {\bibfnamefont {M.}~\bibnamefont
  {{Habouzit}}},\ and\ \bibinfo {author} {\bibfnamefont {M.}~\bibnamefont
  {{Colpi}}},\ }\href {https://doi.org/10.1038/s42254-021-00364-9} {\bibfield
  {journal} {\bibinfo  {journal} {Nature Reviews Physics}\ }\textbf {\bibinfo
  {volume} {3}},\ \bibinfo {pages} {732} (\bibinfo {year} {2021})},\ \Eprint
  {https://arxiv.org/abs/2110.10175} {arXiv:2110.10175 [astro-ph.GA]}
  \BibitemShut {NoStop}%
\bibitem [{\citenamefont {{Rees}}(1984)}]{1984ARA&A..22..471R}%
  \BibitemOpen
  \bibfield  {author} {\bibinfo {author} {\bibfnamefont {M.~J.}\ \bibnamefont
  {{Rees}}},\ }\href {https://doi.org/10.1146/annurev.aa.22.090184.002351}
  {\bibfield  {journal} {\bibinfo  {journal} {\araa}\ }\textbf {\bibinfo
  {volume} {22}},\ \bibinfo {pages} {471} (\bibinfo {year} {1984})}\BibitemShut
  {NoStop}%
\bibitem [{\citenamefont {{Hills}}(1975)}]{1975Natur.254..295H}%
  \BibitemOpen
  \bibfield  {author} {\bibinfo {author} {\bibfnamefont {J.~G.}\ \bibnamefont
  {{Hills}}},\ }\href {https://doi.org/10.1038/254295a0} {\bibfield  {journal}
  {\bibinfo  {journal} {\nat}\ }\textbf {\bibinfo {volume} {254}},\ \bibinfo
  {pages} {295} (\bibinfo {year} {1975})}\BibitemShut {NoStop}%
\bibitem [{\citenamefont {{Rossi}}\ \emph {et~al.}(2021)\citenamefont
  {{Rossi}}, \citenamefont {{Stone}}, \citenamefont {{Law-Smith}},
  \citenamefont {{Macleod}}, \citenamefont {{Lodato}}, \citenamefont {{Dai}},\
  and\ \citenamefont {{Mandel}}}]{Rossi}%
  \BibitemOpen
  \bibfield  {author} {\bibinfo {author} {\bibfnamefont {E.~M.}\ \bibnamefont
  {{Rossi}}}, \bibinfo {author} {\bibfnamefont {N.~C.}\ \bibnamefont
  {{Stone}}}, \bibinfo {author} {\bibfnamefont {J.~A.~P.}\ \bibnamefont
  {{Law-Smith}}}, \bibinfo {author} {\bibfnamefont {M.}~\bibnamefont
  {{Macleod}}}, \bibinfo {author} {\bibfnamefont {G.}~\bibnamefont {{Lodato}}},
  \bibinfo {author} {\bibfnamefont {J.~L.}\ \bibnamefont {{Dai}}},\ and\
  \bibinfo {author} {\bibfnamefont {I.}~\bibnamefont {{Mandel}}},\ }\href
  {https://doi.org/10.1007/s11214-021-00818-7} {\bibfield  {journal} {\bibinfo
  {journal} {\ssr}\ }\textbf {\bibinfo {volume} {217}},\ \bibinfo {eid} {40}
  (\bibinfo {year} {2021})},\ \Eprint {https://arxiv.org/abs/2005.12528}
  {arXiv:2005.12528 [astro-ph.HE]} \BibitemShut {NoStop}%
\bibitem [{\citenamefont {{Stone}}\ \emph {et~al.}(2013)\citenamefont
  {{Stone}}, \citenamefont {{Sari}},\ and\ \citenamefont
  {{Loeb}}}]{2013MNRAS.435.1809S}%
  \BibitemOpen
  \bibfield  {author} {\bibinfo {author} {\bibfnamefont {N.}~\bibnamefont
  {{Stone}}}, \bibinfo {author} {\bibfnamefont {R.}~\bibnamefont {{Sari}}},\
  and\ \bibinfo {author} {\bibfnamefont {A.}~\bibnamefont {{Loeb}}},\ }\href
  {https://doi.org/10.1093/mnras/stt1270} {\bibfield  {journal} {\bibinfo
  {journal} {\mnras}\ }\textbf {\bibinfo {volume} {435}},\ \bibinfo {pages}
  {1809} (\bibinfo {year} {2013})},\ \Eprint {https://arxiv.org/abs/1210.3374}
  {arXiv:1210.3374 [astro-ph.HE]} \BibitemShut {NoStop}%
\bibitem [{\citenamefont {{Shakura}}\ and\ \citenamefont
  {{Sunyaev}}(1973)}]{Shakura1973a}%
  \BibitemOpen
  \bibfield  {author} {\bibinfo {author} {\bibfnamefont {N.~I.}\ \bibnamefont
  {{Shakura}}}\ and\ \bibinfo {author} {\bibfnamefont {R.~A.}\ \bibnamefont
  {{Sunyaev}}},\ }\href@noop {} {\bibfield  {journal} {\bibinfo  {journal}
  {\aap}\ }\textbf {\bibinfo {volume} {500}},\ \bibinfo {pages} {33} (\bibinfo
  {year} {1973})}\BibitemShut {NoStop}%
\bibitem [{\citenamefont {Shibata}\ \emph {et~al.}(2017)\citenamefont
  {Shibata}, \citenamefont {Kiuchi},\ and\ \citenamefont
  {Sekiguchi}}]{Shibata:2017jyf}%
  \BibitemOpen
  \bibfield  {author} {\bibinfo {author} {\bibfnamefont {M.}~\bibnamefont
  {Shibata}}, \bibinfo {author} {\bibfnamefont {K.}~\bibnamefont {Kiuchi}},\
  and\ \bibinfo {author} {\bibfnamefont {Y.-i.}\ \bibnamefont {Sekiguchi}},\
  }\href {https://doi.org/10.1103/PhysRevD.95.083005} {\bibfield  {journal}
  {\bibinfo  {journal} {Phys. Rev. D}\ }\textbf {\bibinfo {volume} {95}},\
  \bibinfo {pages} {083005} (\bibinfo {year} {2017})},\ \Eprint
  {https://arxiv.org/abs/1703.10303} {arXiv:1703.10303 [astro-ph.HE]}
  \BibitemShut {NoStop}%
\bibitem [{\citenamefont {{Font}}\ \emph {et~al.}(1998)\citenamefont {{Font}},
  \citenamefont {{Ib{\'a}{\~n}ez}},\ and\ \citenamefont
  {{Papadopoulos}}}]{1998ApJ...507L..67F}%
  \BibitemOpen
  \bibfield  {author} {\bibinfo {author} {\bibfnamefont {J.~A.}\ \bibnamefont
  {{Font}}}, \bibinfo {author} {\bibfnamefont {J.~M.}\ \bibnamefont
  {{Ib{\'a}{\~n}ez}}},\ and\ \bibinfo {author} {\bibfnamefont {P.}~\bibnamefont
  {{Papadopoulos}}},\ }\href {https://doi.org/10.1086/311666} {\bibfield
  {journal} {\bibinfo  {journal} {\apjl}\ }\textbf {\bibinfo {volume} {507}},\
  \bibinfo {pages} {L67} (\bibinfo {year} {1998})},\ \Eprint
  {https://arxiv.org/abs/astro-ph/9805269} {arXiv:astro-ph/9805269 [astro-ph]}
  \BibitemShut {NoStop}%
\bibitem [{\citenamefont {{Font}}\ \emph {et~al.}(1999)\citenamefont {{Font}},
  \citenamefont {{Ib{\'a}{\~n}ez}},\ and\ \citenamefont
  {{Papadopoulos}}}]{1999MNRAS.305..920F}%
  \BibitemOpen
  \bibfield  {author} {\bibinfo {author} {\bibfnamefont {J.~A.}\ \bibnamefont
  {{Font}}}, \bibinfo {author} {\bibfnamefont {J.~M.}\ \bibnamefont
  {{Ib{\'a}{\~n}ez}}},\ and\ \bibinfo {author} {\bibfnamefont {P.}~\bibnamefont
  {{Papadopoulos}}},\ }\href {https://doi.org/10.1046/j.1365-8711.1999.02459.x}
  {\bibfield  {journal} {\bibinfo  {journal} {\mnras}\ }\textbf {\bibinfo
  {volume} {305}},\ \bibinfo {pages} {920} (\bibinfo {year} {1999})},\ \Eprint
  {https://arxiv.org/abs/astro-ph/9810344} {arXiv:astro-ph/9810344 [astro-ph]}
  \BibitemShut {NoStop}%
\bibitem [{\citenamefont {McKinney}\ and\ \citenamefont
  {Gammie}(2004)}]{McKinney:2004ka}%
  \BibitemOpen
  \bibfield  {author} {\bibinfo {author} {\bibfnamefont {J.~C.}\ \bibnamefont
  {McKinney}}\ and\ \bibinfo {author} {\bibfnamefont {C.~F.}\ \bibnamefont
  {Gammie}},\ }\href {https://doi.org/10.1086/422244} {\bibfield  {journal}
  {\bibinfo  {journal} {Astrophys. J.}\ }\textbf {\bibinfo {volume} {611}},\
  \bibinfo {pages} {977} (\bibinfo {year} {2004})},\ \Eprint
  {https://arxiv.org/abs/astro-ph/0404512} {arXiv:astro-ph/0404512}
  \BibitemShut {NoStop}%
\bibitem [{\citenamefont {Israel}\ and\ \citenamefont
  {Stewart}(1979)}]{Israel:1979wp}%
  \BibitemOpen
  \bibfield  {author} {\bibinfo {author} {\bibfnamefont {W.}~\bibnamefont
  {Israel}}\ and\ \bibinfo {author} {\bibfnamefont {J.~M.}\ \bibnamefont
  {Stewart}},\ }\href {https://doi.org/10.1016/0003-4916(79)90130-1} {\bibfield
   {journal} {\bibinfo  {journal} {Annals Phys.}\ }\textbf {\bibinfo {volume}
  {118}},\ \bibinfo {pages} {341} (\bibinfo {year} {1979})}\BibitemShut
  {NoStop}%
\bibitem [{\citenamefont {Balbus}\ and\ \citenamefont
  {Hawley}(1998)}]{Balbus:1998ja}%
  \BibitemOpen
  \bibfield  {author} {\bibinfo {author} {\bibfnamefont {S.~A.}\ \bibnamefont
  {Balbus}}\ and\ \bibinfo {author} {\bibfnamefont {J.~F.}\ \bibnamefont
  {Hawley}},\ }\href {https://doi.org/10.1103/RevModPhys.70.1} {\bibfield
  {journal} {\bibinfo  {journal} {Rev. Mod. Phys.}\ }\textbf {\bibinfo {volume}
  {70}},\ \bibinfo {pages} {1} (\bibinfo {year} {1998})}\BibitemShut {NoStop}%
\bibitem [{\citenamefont {Suzuki}\ and\ \citenamefont
  {Inutsuka}(2014)}]{Suzuki:2013rka}%
  \BibitemOpen
  \bibfield  {author} {\bibinfo {author} {\bibfnamefont {T.~K.}\ \bibnamefont
  {Suzuki}}\ and\ \bibinfo {author} {\bibfnamefont {S.-i.}\ \bibnamefont
  {Inutsuka}},\ }\href {https://doi.org/10.1088/0004-637X/784/2/121} {\bibfield
   {journal} {\bibinfo  {journal} {Astrophys. J.}\ }\textbf {\bibinfo {volume}
  {784}},\ \bibinfo {pages} {121} (\bibinfo {year} {2014})},\ \Eprint
  {https://arxiv.org/abs/1309.6916} {arXiv:1309.6916 [astro-ph.EP]}
  \BibitemShut {NoStop}%
\bibitem [{\citenamefont {Shi}\ \emph {et~al.}(2016)\citenamefont {Shi},
  \citenamefont {Stone},\ and\ \citenamefont {Huang}}]{Shi:2015mvh}%
  \BibitemOpen
  \bibfield  {author} {\bibinfo {author} {\bibfnamefont {J.-M.}\ \bibnamefont
  {Shi}}, \bibinfo {author} {\bibfnamefont {J.~M.}\ \bibnamefont {Stone}},\
  and\ \bibinfo {author} {\bibfnamefont {C.~X.}\ \bibnamefont {Huang}},\ }\href
  {https://doi.org/10.1093/mnras/stv2815} {\bibfield  {journal} {\bibinfo
  {journal} {Mon. Not. Roy. Astron. Soc.}\ }\textbf {\bibinfo {volume} {456}},\
  \bibinfo {pages} {2273} (\bibinfo {year} {2016})},\ \Eprint
  {https://arxiv.org/abs/1512.01106} {arXiv:1512.01106 [astro-ph.HE]}
  \BibitemShut {NoStop}%
\bibitem [{\citenamefont {{Fishbone}}\ and\ \citenamefont
  {{Moncrief}}(1976)}]{1976ApJ...207..962F}%
  \BibitemOpen
  \bibfield  {author} {\bibinfo {author} {\bibfnamefont {L.~G.}\ \bibnamefont
  {{Fishbone}}}\ and\ \bibinfo {author} {\bibfnamefont {V.}~\bibnamefont
  {{Moncrief}}},\ }\href {https://doi.org/10.1086/154565} {\bibfield  {journal}
  {\bibinfo  {journal} {\apj}\ }\textbf {\bibinfo {volume} {207}},\ \bibinfo
  {pages} {962} (\bibinfo {year} {1976})}\BibitemShut {NoStop}%
\bibitem [{\citenamefont {{Papaloizou}}\ and\ \citenamefont
  {{Pringle}}(1984)}]{1984MNRAS.208..721P}%
  \BibitemOpen
  \bibfield  {author} {\bibinfo {author} {\bibfnamefont {J.~C.~B.}\
  \bibnamefont {{Papaloizou}}}\ and\ \bibinfo {author} {\bibfnamefont {J.~E.}\
  \bibnamefont {{Pringle}}},\ }\href {https://doi.org/10.1093/mnras/208.4.721}
  {\bibfield  {journal} {\bibinfo  {journal} {\mnras}\ }\textbf {\bibinfo
  {volume} {208}},\ \bibinfo {pages} {721} (\bibinfo {year}
  {1984})}\BibitemShut {NoStop}%
\bibitem [{\citenamefont {{Kojima}}(1986)}]{1986PThPh..75.1464K}%
  \BibitemOpen
  \bibfield  {author} {\bibinfo {author} {\bibfnamefont {Y.}~\bibnamefont
  {{Kojima}}},\ }\href {https://doi.org/10.1143/PTP.75.1464} {\bibfield
  {journal} {\bibinfo  {journal} {Progress of Theoretical Physics}\ }\textbf
  {\bibinfo {volume} {75}},\ \bibinfo {pages} {1464} (\bibinfo {year}
  {1986})}\BibitemShut {NoStop}%
\bibitem [{\citenamefont {Lam}\ and\ \citenamefont
  {Shibata}(2025)}]{Lam:2025pmz}%
  \BibitemOpen
  \bibfield  {author} {\bibinfo {author} {\bibfnamefont {A.~T.-L.}\
  \bibnamefont {Lam}}\ and\ \bibinfo {author} {\bibfnamefont {M.}~\bibnamefont
  {Shibata}},\ }\href@noop {} {\bibfield  {journal} {\bibinfo  {journal} {arXiv
  e-prints}\ } (\bibinfo {year} {2025})},\ \Eprint
  {https://arxiv.org/abs/2502.03223} {arXiv:2502.03223 [astro-ph.HE]}
  \BibitemShut {NoStop}%
\bibitem [{\citenamefont {{Gammie}}\ \emph {et~al.}(2004)\citenamefont
  {{Gammie}}, \citenamefont {{Shapiro}},\ and\ \citenamefont
  {{McKinney}}}]{2004ApJ...602..312G}%
  \BibitemOpen
  \bibfield  {author} {\bibinfo {author} {\bibfnamefont {C.~F.}\ \bibnamefont
  {{Gammie}}}, \bibinfo {author} {\bibfnamefont {S.~L.}\ \bibnamefont
  {{Shapiro}}},\ and\ \bibinfo {author} {\bibfnamefont {J.~C.}\ \bibnamefont
  {{McKinney}}},\ }\href {https://doi.org/10.1086/380996} {\bibfield  {journal}
  {\bibinfo  {journal} {\apj}\ }\textbf {\bibinfo {volume} {602}},\ \bibinfo
  {pages} {312} (\bibinfo {year} {2004})},\ \Eprint
  {https://arxiv.org/abs/astro-ph/0310886} {arXiv:astro-ph/0310886 [astro-ph]}
  \BibitemShut {NoStop}%
\bibitem [{\citenamefont {{Bardeen}}\ \emph {et~al.}(1972)\citenamefont
  {{Bardeen}}, \citenamefont {{Press}},\ and\ \citenamefont
  {{Teukolsky}}}]{1972ApJ...178..347B}%
  \BibitemOpen
  \bibfield  {author} {\bibinfo {author} {\bibfnamefont {J.~M.}\ \bibnamefont
  {{Bardeen}}}, \bibinfo {author} {\bibfnamefont {W.~H.}\ \bibnamefont
  {{Press}}},\ and\ \bibinfo {author} {\bibfnamefont {S.~A.}\ \bibnamefont
  {{Teukolsky}}},\ }\href {https://doi.org/10.1086/151796} {\bibfield
  {journal} {\bibinfo  {journal} {\apj}\ }\textbf {\bibinfo {volume} {178}},\
  \bibinfo {pages} {347} (\bibinfo {year} {1972})}\BibitemShut {NoStop}%
\bibitem [{\citenamefont {{Shapiro}}\ and\ \citenamefont
  {{Teukolsky}}(1983)}]{1983bhwd.book.....S}%
  \BibitemOpen
  \bibfield  {author} {\bibinfo {author} {\bibfnamefont {S.~L.}\ \bibnamefont
  {{Shapiro}}}\ and\ \bibinfo {author} {\bibfnamefont {S.~A.}\ \bibnamefont
  {{Teukolsky}}},\ }\href {https://doi.org/10.1002/9783527617661} {\emph
  {\bibinfo {title} {{Black holes, white dwarfs and neutron stars. The physics
  of compact objects}}}}\ (\bibinfo  {publisher} {John Wiley \& Sons},\
  \bibinfo {year} {1983})\BibitemShut {NoStop}%
\bibitem [{\citenamefont {{Fujibayashi}}\ \emph {et~al.}(2020)\citenamefont
  {{Fujibayashi}}, \citenamefont {{Shibata}}, \citenamefont {{Wanajo}},
  \citenamefont {{Kiuchi}}, \citenamefont {{Kyutoku}},\ and\ \citenamefont
  {{Sekiguchi}}}]{Fujibayashi2020a}%
  \BibitemOpen
  \bibfield  {author} {\bibinfo {author} {\bibfnamefont {S.}~\bibnamefont
  {{Fujibayashi}}}, \bibinfo {author} {\bibfnamefont {M.}~\bibnamefont
  {{Shibata}}}, \bibinfo {author} {\bibfnamefont {S.}~\bibnamefont {{Wanajo}}},
  \bibinfo {author} {\bibfnamefont {K.}~\bibnamefont {{Kiuchi}}}, \bibinfo
  {author} {\bibfnamefont {K.}~\bibnamefont {{Kyutoku}}},\ and\ \bibinfo
  {author} {\bibfnamefont {Y.}~\bibnamefont {{Sekiguchi}}},\ }\href
  {https://doi.org/10.1103/PhysRevD.101.083029} {\bibfield  {journal} {\bibinfo
   {journal} {\prd}\ }\textbf {\bibinfo {volume} {101}},\ \bibinfo {eid}
  {083029} (\bibinfo {year} {2020})},\ \Eprint
  {https://arxiv.org/abs/2001.04467} {arXiv:2001.04467 [astro-ph.HE]}
  \BibitemShut {NoStop}%
\bibitem [{\citenamefont {{Sadowski}}\ \emph {et~al.}(2014)\citenamefont
  {{Sadowski}}, \citenamefont {{Narayan}}, \citenamefont {{McKinney}},\ and\
  \citenamefont {{Tchekhovskoy}}}]{2014MNRAS.439..503S}%
  \BibitemOpen
  \bibfield  {author} {\bibinfo {author} {\bibfnamefont {A.}~\bibnamefont
  {{Sadowski}}}, \bibinfo {author} {\bibfnamefont {R.}~\bibnamefont
  {{Narayan}}}, \bibinfo {author} {\bibfnamefont {J.~C.}\ \bibnamefont
  {{McKinney}}},\ and\ \bibinfo {author} {\bibfnamefont {A.}~\bibnamefont
  {{Tchekhovskoy}}},\ }\href {https://doi.org/10.1093/mnras/stt2479} {\bibfield
   {journal} {\bibinfo  {journal} {\mnras}\ }\textbf {\bibinfo {volume}
  {439}},\ \bibinfo {pages} {503} (\bibinfo {year} {2014})},\ \Eprint
  {https://arxiv.org/abs/1311.5900} {arXiv:1311.5900 [astro-ph.HE]}
  \BibitemShut {NoStop}%
\bibitem [{\citenamefont {{Jiang}}\ \emph {et~al.}(2014)\citenamefont
  {{Jiang}}, \citenamefont {{Stone}},\ and\ \citenamefont
  {{Davis}}}]{2014ApJ...796..106J}%
  \BibitemOpen
  \bibfield  {author} {\bibinfo {author} {\bibfnamefont {Y.-F.}\ \bibnamefont
  {{Jiang}}}, \bibinfo {author} {\bibfnamefont {J.~M.}\ \bibnamefont
  {{Stone}}},\ and\ \bibinfo {author} {\bibfnamefont {S.~W.}\ \bibnamefont
  {{Davis}}},\ }\href {https://doi.org/10.1088/0004-637X/796/2/106} {\bibfield
  {journal} {\bibinfo  {journal} {\apj}\ }\textbf {\bibinfo {volume} {796}},\
  \bibinfo {eid} {106} (\bibinfo {year} {2014})},\ \Eprint
  {https://arxiv.org/abs/1410.0678} {arXiv:1410.0678 [astro-ph.HE]}
  \BibitemShut {NoStop}%
\bibitem [{\citenamefont {{Sadowski}}\ \emph {et~al.}(2015)\citenamefont
  {{Sadowski}}, \citenamefont {{Narayan}}, \citenamefont {{Tchekhovskoy}},
  \citenamefont {{Abarca}}, \citenamefont {{Zhu}},\ and\ \citenamefont
  {{McKinney}}}]{2015MNRAS.447...49S}%
  \BibitemOpen
  \bibfield  {author} {\bibinfo {author} {\bibfnamefont {A.}~\bibnamefont
  {{Sadowski}}}, \bibinfo {author} {\bibfnamefont {R.}~\bibnamefont
  {{Narayan}}}, \bibinfo {author} {\bibfnamefont {A.}~\bibnamefont
  {{Tchekhovskoy}}}, \bibinfo {author} {\bibfnamefont {D.}~\bibnamefont
  {{Abarca}}}, \bibinfo {author} {\bibfnamefont {Y.}~\bibnamefont {{Zhu}}},\
  and\ \bibinfo {author} {\bibfnamefont {J.~C.}\ \bibnamefont {{McKinney}}},\
  }\href {https://doi.org/10.1093/mnras/stu2387} {\bibfield  {journal}
  {\bibinfo  {journal} {\mnras}\ }\textbf {\bibinfo {volume} {447}},\ \bibinfo
  {pages} {49} (\bibinfo {year} {2015})},\ \Eprint
  {https://arxiv.org/abs/1407.4421} {arXiv:1407.4421 [astro-ph.HE]}
  \BibitemShut {NoStop}%
\bibitem [{\citenamefont {{Sadowski}}\ and\ \citenamefont
  {{Narayan}}(2016)}]{2016MNRAS.456.3929S}%
  \BibitemOpen
  \bibfield  {author} {\bibinfo {author} {\bibfnamefont {A.}~\bibnamefont
  {{Sadowski}}}\ and\ \bibinfo {author} {\bibfnamefont {R.}~\bibnamefont
  {{Narayan}}},\ }\href {https://doi.org/10.1093/mnras/stv2941} {\bibfield
  {journal} {\bibinfo  {journal} {\mnras}\ }\textbf {\bibinfo {volume} {456}},\
  \bibinfo {pages} {3929} (\bibinfo {year} {2016})},\ \Eprint
  {https://arxiv.org/abs/1509.03168} {arXiv:1509.03168 [astro-ph.HE]}
  \BibitemShut {NoStop}%
\bibitem [{\citenamefont {{Kitaki}}\ \emph {et~al.}(2018)\citenamefont
  {{Kitaki}}, \citenamefont {{Mineshige}}, \citenamefont {{Ohsuga}},\ and\
  \citenamefont {{Kawashima}}}]{2018PASJ...70..108K}%
  \BibitemOpen
  \bibfield  {author} {\bibinfo {author} {\bibfnamefont {T.}~\bibnamefont
  {{Kitaki}}}, \bibinfo {author} {\bibfnamefont {S.}~\bibnamefont
  {{Mineshige}}}, \bibinfo {author} {\bibfnamefont {K.}~\bibnamefont
  {{Ohsuga}}},\ and\ \bibinfo {author} {\bibfnamefont {T.}~\bibnamefont
  {{Kawashima}}},\ }\href {https://doi.org/10.1093/pasj/psy110} {\bibfield
  {journal} {\bibinfo  {journal} {\pasj}\ }\textbf {\bibinfo {volume} {70}},\
  \bibinfo {eid} {108} (\bibinfo {year} {2018})},\ \Eprint
  {https://arxiv.org/abs/1809.01151} {arXiv:1809.01151 [astro-ph.HE]}
  \BibitemShut {NoStop}%
\bibitem [{\citenamefont {{Yoshioka}}\ \emph {et~al.}(2022)\citenamefont
  {{Yoshioka}}, \citenamefont {{Mineshige}}, \citenamefont {{Ohsuga}},
  \citenamefont {{Kawashima}},\ and\ \citenamefont
  {{Kitaki}}}]{2022PASJ...74.1378Y}%
  \BibitemOpen
  \bibfield  {author} {\bibinfo {author} {\bibfnamefont {S.}~\bibnamefont
  {{Yoshioka}}}, \bibinfo {author} {\bibfnamefont {S.}~\bibnamefont
  {{Mineshige}}}, \bibinfo {author} {\bibfnamefont {K.}~\bibnamefont
  {{Ohsuga}}}, \bibinfo {author} {\bibfnamefont {T.}~\bibnamefont
  {{Kawashima}}},\ and\ \bibinfo {author} {\bibfnamefont {T.}~\bibnamefont
  {{Kitaki}}},\ }\href {https://doi.org/10.1093/pasj/psac076} {\bibfield
  {journal} {\bibinfo  {journal} {\pasj}\ }\textbf {\bibinfo {volume} {74}},\
  \bibinfo {pages} {1378} (\bibinfo {year} {2022})},\ \Eprint
  {https://arxiv.org/abs/2209.01427} {arXiv:2209.01427 [astro-ph.HE]}
  \BibitemShut {NoStop}%
\bibitem [{\citenamefont {{Hu}}\ \emph {et~al.}(2022)\citenamefont {{Hu}},
  \citenamefont {{Inayoshi}}, \citenamefont {{Haiman}}, \citenamefont
  {{Quataert}},\ and\ \citenamefont {{Kuiper}}}]{2022ApJ...934..132H}%
  \BibitemOpen
  \bibfield  {author} {\bibinfo {author} {\bibfnamefont {H.}~\bibnamefont
  {{Hu}}}, \bibinfo {author} {\bibfnamefont {K.}~\bibnamefont {{Inayoshi}}},
  \bibinfo {author} {\bibfnamefont {Z.}~\bibnamefont {{Haiman}}}, \bibinfo
  {author} {\bibfnamefont {E.}~\bibnamefont {{Quataert}}},\ and\ \bibinfo
  {author} {\bibfnamefont {R.}~\bibnamefont {{Kuiper}}},\ }\href
  {https://doi.org/10.3847/1538-4357/ac75d8} {\bibfield  {journal} {\bibinfo
  {journal} {\apj}\ }\textbf {\bibinfo {volume} {934}},\ \bibinfo {eid} {132}
  (\bibinfo {year} {2022})},\ \Eprint {https://arxiv.org/abs/2203.14994}
  {arXiv:2203.14994 [astro-ph.HE]} \BibitemShut {NoStop}%
\bibitem [{\citenamefont {{Stone}}\ \emph {et~al.}(1999)\citenamefont
  {{Stone}}, \citenamefont {{Pringle}},\ and\ \citenamefont
  {{Begelman}}}]{1999MNRAS.310.1002S}%
  \BibitemOpen
  \bibfield  {author} {\bibinfo {author} {\bibfnamefont {J.~M.}\ \bibnamefont
  {{Stone}}}, \bibinfo {author} {\bibfnamefont {J.~E.}\ \bibnamefont
  {{Pringle}}},\ and\ \bibinfo {author} {\bibfnamefont {M.~C.}\ \bibnamefont
  {{Begelman}}},\ }\href {https://doi.org/10.1046/j.1365-8711.1999.03024.x}
  {\bibfield  {journal} {\bibinfo  {journal} {\mnras}\ }\textbf {\bibinfo
  {volume} {310}},\ \bibinfo {pages} {1002} (\bibinfo {year} {1999})},\ \Eprint
  {https://arxiv.org/abs/astro-ph/9908185} {arXiv:astro-ph/9908185 [astro-ph]}
  \BibitemShut {NoStop}%
\bibitem [{\citenamefont {{Narayan}}\ \emph {et~al.}(2000)\citenamefont
  {{Narayan}}, \citenamefont {{Igumenshchev}},\ and\ \citenamefont
  {{Abramowicz}}}]{2000ApJ...539..798N}%
  \BibitemOpen
  \bibfield  {author} {\bibinfo {author} {\bibfnamefont {R.}~\bibnamefont
  {{Narayan}}}, \bibinfo {author} {\bibfnamefont {I.~V.}\ \bibnamefont
  {{Igumenshchev}}},\ and\ \bibinfo {author} {\bibfnamefont {M.~A.}\
  \bibnamefont {{Abramowicz}}},\ }\href {https://doi.org/10.1086/309268}
  {\bibfield  {journal} {\bibinfo  {journal} {\apj}\ }\textbf {\bibinfo
  {volume} {539}},\ \bibinfo {pages} {798} (\bibinfo {year} {2000})},\ \Eprint
  {https://arxiv.org/abs/astro-ph/9912449} {arXiv:astro-ph/9912449 [astro-ph]}
  \BibitemShut {NoStop}%
\bibitem [{\citenamefont {{Quataert}}\ and\ \citenamefont
  {{Gruzinov}}(2000)}]{2000ApJ...539..809Q}%
  \BibitemOpen
  \bibfield  {author} {\bibinfo {author} {\bibfnamefont {E.}~\bibnamefont
  {{Quataert}}}\ and\ \bibinfo {author} {\bibfnamefont {A.}~\bibnamefont
  {{Gruzinov}}},\ }\href {https://doi.org/10.1086/309267} {\bibfield  {journal}
  {\bibinfo  {journal} {\apj}\ }\textbf {\bibinfo {volume} {539}},\ \bibinfo
  {pages} {809} (\bibinfo {year} {2000})},\ \Eprint
  {https://arxiv.org/abs/astro-ph/9912440} {arXiv:astro-ph/9912440 [astro-ph]}
  \BibitemShut {NoStop}%
\bibitem [{\citenamefont {{Yuan}}\ and\ \citenamefont
  {{Narayan}}(2014)}]{2014ARA&A..52..529Y}%
  \BibitemOpen
  \bibfield  {author} {\bibinfo {author} {\bibfnamefont {F.}~\bibnamefont
  {{Yuan}}}\ and\ \bibinfo {author} {\bibfnamefont {R.}~\bibnamefont
  {{Narayan}}},\ }\href {https://doi.org/10.1146/annurev-astro-082812-141003}
  {\bibfield  {journal} {\bibinfo  {journal} {\araa}\ }\textbf {\bibinfo
  {volume} {52}},\ \bibinfo {pages} {529} (\bibinfo {year} {2014})},\ \Eprint
  {https://arxiv.org/abs/1401.0586} {arXiv:1401.0586 [astro-ph.HE]}
  \BibitemShut {NoStop}%
\bibitem [{\citenamefont {{Toyouchi}}\ \emph {et~al.}(2024)\citenamefont
  {{Toyouchi}}, \citenamefont {{Hotokezaka}}, \citenamefont {{Inayoshi}},\ and\
  \citenamefont {{Kuiper}}}]{2024MNRAS.532.4826T}%
  \BibitemOpen
  \bibfield  {author} {\bibinfo {author} {\bibfnamefont {D.}~\bibnamefont
  {{Toyouchi}}}, \bibinfo {author} {\bibfnamefont {K.}~\bibnamefont
  {{Hotokezaka}}}, \bibinfo {author} {\bibfnamefont {K.}~\bibnamefont
  {{Inayoshi}}},\ and\ \bibinfo {author} {\bibfnamefont {R.}~\bibnamefont
  {{Kuiper}}},\ }\href {https://doi.org/10.1093/mnras/stae1798} {\bibfield
  {journal} {\bibinfo  {journal} {\mnras}\ }\textbf {\bibinfo {volume} {532}},\
  \bibinfo {pages} {4826} (\bibinfo {year} {2024})},\ \Eprint
  {https://arxiv.org/abs/2405.07190} {arXiv:2405.07190 [astro-ph.HE]}
  \BibitemShut {NoStop}%
\bibitem [{\citenamefont {Parker}\ \emph {et~al.}(2017)\citenamefont {Parker}
  \emph {et~al.}}]{Parker:2017wnh}%
  \BibitemOpen
  \bibfield  {author} {\bibinfo {author} {\bibfnamefont {M.~L.}\ \bibnamefont
  {Parker}} \emph {et~al.},\ }\href {https://doi.org/10.1038/nature21385}
  {\bibfield  {journal} {\bibinfo  {journal} {Nature}\ }\textbf {\bibinfo
  {volume} {543}},\ \bibinfo {pages} {83} (\bibinfo {year} {2017})},\ \Eprint
  {https://arxiv.org/abs/1703.00071} {arXiv:1703.00071 [astro-ph.HE]}
  \BibitemShut {NoStop}%
\bibitem [{\citenamefont {{Narayan}}\ \emph {et~al.}(2022)\citenamefont
  {{Narayan}}, \citenamefont {{Chael}}, \citenamefont {{Chatterjee}},
  \citenamefont {{Ricarte}},\ and\ \citenamefont
  {{Curd}}}]{2022MNRAS.511.3795N}%
  \BibitemOpen
  \bibfield  {author} {\bibinfo {author} {\bibfnamefont {R.}~\bibnamefont
  {{Narayan}}}, \bibinfo {author} {\bibfnamefont {A.}~\bibnamefont {{Chael}}},
  \bibinfo {author} {\bibfnamefont {K.}~\bibnamefont {{Chatterjee}}}, \bibinfo
  {author} {\bibfnamefont {A.}~\bibnamefont {{Ricarte}}},\ and\ \bibinfo
  {author} {\bibfnamefont {B.}~\bibnamefont {{Curd}}},\ }\href
  {https://doi.org/10.1093/mnras/stac285} {\bibfield  {journal} {\bibinfo
  {journal} {\mnras}\ }\textbf {\bibinfo {volume} {511}},\ \bibinfo {pages}
  {3795} (\bibinfo {year} {2022})},\ \Eprint {https://arxiv.org/abs/2108.12380}
  {arXiv:2108.12380 [astro-ph.HE]} \BibitemShut {NoStop}%
\bibitem [{\citenamefont {{Blandford}}\ and\ \citenamefont
  {{Znajek}}(1977)}]{Blandford1977}%
  \BibitemOpen
  \bibfield  {author} {\bibinfo {author} {\bibfnamefont {R.~D.}\ \bibnamefont
  {{Blandford}}}\ and\ \bibinfo {author} {\bibfnamefont {R.~L.}\ \bibnamefont
  {{Znajek}}},\ }\href {https://doi.org/10.1093/mnras/179.3.433} {\bibfield
  {journal} {\bibinfo  {journal} {\mnras}\ }\textbf {\bibinfo {volume} {179}},\
  \bibinfo {pages} {433} (\bibinfo {year} {1977})}\BibitemShut {NoStop}%
\bibitem [{\citenamefont {{Syer}}\ and\ \citenamefont
  {{Ulmer}}(1999)}]{1999MNRAS.306...35S}%
  \BibitemOpen
  \bibfield  {author} {\bibinfo {author} {\bibfnamefont {D.}~\bibnamefont
  {{Syer}}}\ and\ \bibinfo {author} {\bibfnamefont {A.}~\bibnamefont
  {{Ulmer}}},\ }\href {https://doi.org/10.1046/j.1365-8711.1999.02445.x}
  {\bibfield  {journal} {\bibinfo  {journal} {\mnras}\ }\textbf {\bibinfo
  {volume} {306}},\ \bibinfo {pages} {35} (\bibinfo {year} {1999})},\ \Eprint
  {https://arxiv.org/abs/astro-ph/9812389} {arXiv:astro-ph/9812389 [astro-ph]}
  \BibitemShut {NoStop}%
\bibitem [{\citenamefont {{Magorrian}}\ and\ \citenamefont
  {{Tremaine}}(1999)}]{1999MNRAS.309..447M}%
  \BibitemOpen
  \bibfield  {author} {\bibinfo {author} {\bibfnamefont {J.}~\bibnamefont
  {{Magorrian}}}\ and\ \bibinfo {author} {\bibfnamefont {S.}~\bibnamefont
  {{Tremaine}}},\ }\href {https://doi.org/10.1046/j.1365-8711.1999.02853.x}
  {\bibfield  {journal} {\bibinfo  {journal} {\mnras}\ }\textbf {\bibinfo
  {volume} {309}},\ \bibinfo {pages} {447} (\bibinfo {year} {1999})},\ \Eprint
  {https://arxiv.org/abs/astro-ph/9902032} {arXiv:astro-ph/9902032 [astro-ph]}
  \BibitemShut {NoStop}%
\bibitem [{\citenamefont {{Hosokawa}}\ \emph {et~al.}(2013)\citenamefont
  {{Hosokawa}}, \citenamefont {{Yorke}}, \citenamefont {{Inayoshi}},
  \citenamefont {{Omukai}},\ and\ \citenamefont
  {{Yoshida}}}]{2013ApJ...778..178H}%
  \BibitemOpen
  \bibfield  {author} {\bibinfo {author} {\bibfnamefont {T.}~\bibnamefont
  {{Hosokawa}}}, \bibinfo {author} {\bibfnamefont {H.~W.}\ \bibnamefont
  {{Yorke}}}, \bibinfo {author} {\bibfnamefont {K.}~\bibnamefont {{Inayoshi}}},
  \bibinfo {author} {\bibfnamefont {K.}~\bibnamefont {{Omukai}}},\ and\
  \bibinfo {author} {\bibfnamefont {N.}~\bibnamefont {{Yoshida}}},\ }\href
  {https://doi.org/10.1088/0004-637X/778/2/178} {\bibfield  {journal} {\bibinfo
   {journal} {\apj}\ }\textbf {\bibinfo {volume} {778}},\ \bibinfo {eid} {178}
  (\bibinfo {year} {2013})},\ \Eprint {https://arxiv.org/abs/1308.4457}
  {arXiv:1308.4457 [astro-ph.SR]} \BibitemShut {NoStop}%
\bibitem [{\citenamefont {{Sarazin}}(1986)}]{1986RvMP...58....1S}%
  \BibitemOpen
  \bibfield  {author} {\bibinfo {author} {\bibfnamefont {C.~L.}\ \bibnamefont
  {{Sarazin}}},\ }\href {https://doi.org/10.1103/RevModPhys.58.1} {\bibfield
  {journal} {\bibinfo  {journal} {Reviews of Modern Physics}\ }\textbf
  {\bibinfo {volume} {58}},\ \bibinfo {pages} {1} (\bibinfo {year}
  {1986})}\BibitemShut {NoStop}%
\bibitem [{\citenamefont {{McKinney}}\ \emph {et~al.}(2014)\citenamefont
  {{McKinney}}, \citenamefont {{Tchekhovskoy}}, \citenamefont {{Sadowski}},\
  and\ \citenamefont {{Narayan}}}]{2014MNRAS.441.3177M}%
  \BibitemOpen
  \bibfield  {author} {\bibinfo {author} {\bibfnamefont {J.~C.}\ \bibnamefont
  {{McKinney}}}, \bibinfo {author} {\bibfnamefont {A.}~\bibnamefont
  {{Tchekhovskoy}}}, \bibinfo {author} {\bibfnamefont {A.}~\bibnamefont
  {{Sadowski}}},\ and\ \bibinfo {author} {\bibfnamefont {R.}~\bibnamefont
  {{Narayan}}},\ }\href {https://doi.org/10.1093/mnras/stu762} {\bibfield
  {journal} {\bibinfo  {journal} {\mnras}\ }\textbf {\bibinfo {volume} {441}},\
  \bibinfo {pages} {3177} (\bibinfo {year} {2014})},\ \Eprint
  {https://arxiv.org/abs/1312.6127} {arXiv:1312.6127 [astro-ph.CO]}
  \BibitemShut {NoStop}%
\bibitem [{\citenamefont {{Abramowicz}}\ \emph {et~al.}(2002)\citenamefont
  {{Abramowicz}}, \citenamefont {{Igumenshchev}}, \citenamefont {{Quataert}},\
  and\ \citenamefont {{Narayan}}}]{2002ApJ...565.1101A}%
  \BibitemOpen
  \bibfield  {author} {\bibinfo {author} {\bibfnamefont {M.~A.}\ \bibnamefont
  {{Abramowicz}}}, \bibinfo {author} {\bibfnamefont {I.~V.}\ \bibnamefont
  {{Igumenshchev}}}, \bibinfo {author} {\bibfnamefont {E.}~\bibnamefont
  {{Quataert}}},\ and\ \bibinfo {author} {\bibfnamefont {R.}~\bibnamefont
  {{Narayan}}},\ }\href {https://doi.org/10.1086/324717} {\bibfield  {journal}
  {\bibinfo  {journal} {\apj}\ }\textbf {\bibinfo {volume} {565}},\ \bibinfo
  {pages} {1101} (\bibinfo {year} {2002})},\ \Eprint
  {https://arxiv.org/abs/astro-ph/0110371} {arXiv:astro-ph/0110371 [astro-ph]}
  \BibitemShut {NoStop}%
\end{thebibliography}%

\end{document}